\documentclass[useAMS,usenatbib]{mn2e}
\usepackage{epsfig}
\usepackage{rotating}
\usepackage{natbib}
\usepackage{comment}
\def\etal{{et al.\thinspace}}

\def\spose#1{\hbox to 0pt{#1\hss}}

\def\multleft#1{\hbox to size{\vbox {\halign {\lft{##}\cr #1}}\hfill}\par}
\def\multright#1{\hbox to size{\vbox {\halign {\rt{##}\cr #1}}\hfill}\par}

\def\degmark{^\circ}
\def\boxit#1{\vbox{\hrule\hbox{\vrule\kern3pt\vbox{\kern3pt
          #1 \kern3pt}\kern3pt\vrule}\hrule}}

\def\cm{{\rm\thinspace cm}}

\def\erg{{\rm\thinspace erg}}
\def\eV{{\rm\thinspace eV}}

\def\keV{{\rm\thinspace keV}}
\def\km{{\rm\thinspace km}}

\def\Mpc{{\rm\thinspace Mpc}}
\def\Msun{\hbox{$\rm\thinspace M_{\odot}$}}

\def\ph{{\rm\thinspace ph}}
\def\s{{\rm\thinspace s}}
\def\ks{{\rm\thinspace ks}}

\def\cts{{\rm\thinspace cts}}

\def\pcmcu{\hbox{$\cm^{-3}\,$}}

\def\ergcmps{\hbox{$\erg\cm\s^{-1}\,$}}

\def\ergpcmsqps{\hbox{$\erg\cm^{-2}\s^{-1}\,$}}

\def\ergps{\hbox{$\erg\s^{-1}\,$}}

\def\kmps{\hbox{$\km\s^{-1}\,$}}

\def\pcmsq{\hbox{$\cm^{-2}\,$}}
\def\pcmcu{\hbox{$\cm^{-3}\,$}}

\def\phpcmsqps{\hbox{$\ph\cm^{-2}\s^{-1}\,$}}

\def\kmpspMpc{\hbox{$\kmps\Mpc^{-1}$}}

\def\ctsps{\hbox{$\cts\s^{-1}$}}

\makeatletter

\let\@internalcite\cite
\def\cite{\@ifstar{\citey}{\citefull}}
\def\citefull{\def\astroncite##1##2{##1\ ##2}\@internalcite}
\def\citey{\def\astroncite##1##2{##1\ (##2)}\@internalcite}
\def\citeyear{\def\astroncite##1##2{##2}\@internalcite}
\def\citename{\def\astroncite##1##2{##1}\@internalcite}
\def\@citex[#1]#2{\if@filesw\immediate\write\@auxout{\string\citation{#2}}\fi
  \def\@citea{}\@cite{\@for\@citeb:=#2\do
    {\@citea\def\@citea{; }\@ifundefined
       {b@\@citeb}{{\bf ??}\@warning
       {Citation `\@citeb' on page \thepage \space undefined}}%
{\csname b@\@citeb\endcsname}}}{#1}}

\def\@cite#1#2{#1\if@tempswa #2\fi}
\def\@biblabel#1{}

\def\astroncite#1#2{#1\ #2}
\makeatother

\title[An Examination of the Spectral Variability in NGC~1365 with
  {\em Suzaku}]{An Examination of the Spectral Variability in NGC~1365
  with {\em Suzaku}}
\author[L.~W.~Brenneman \etal]{L.~W.~Brenneman$^{1}$\thanks{E-mail:
  lbrenneman@cfa.harvard.edu}, G.~Risaliti$^{1,2}$, M.~Elvis$^{1}$,
and E.~Nardini$^{1}$\\
$^{1}$Harvard-Smithsonian Center for Astrophysics, 60 Garden St., Cambridge, MA~02138~USA\\
$^{2}$INAF - Osservatorio Astrofisico di Arcetri, Largo E. Fermi 5, I-50125 Firenze, Italy}

\begin{document}

\date{In original form 2012 September 19}

\pagerange{\pageref{firstpage}--\pageref{lastpage}} \pubyear{2012}

\maketitle

\label{firstpage}

\begin{abstract}
We present jointly analyzed data from three deep {\it Suzaku}
observations of NGC~1365.  These high signal-to-noise spectra enable
us to examine the nature of this variable, obscured AGN in unprecedented
detail on timescales ranging from hours to years.  We find that, in
addition to the power-law continuum and absorption from ionized gas
seen in most AGN, inner disk reflection and variable absorption from
neutral gas within the Broad Emission Line Region are both necessary
components in all three observations.
We confirm the clumpy nature of the cold absorbing gas, though we
note that occultations of the inner disk and corona are much more
pronounced in the high-flux
state (2008) than in the low-flux state (2010) of the source.  The
onset and duration of the ``dips'' in the X-ray light curve in 2010
are both significantly longer than in 2008, however, indicating that
either the distance to the gas from the black hole is larger, or that the
nature of the gas has changed between epochs.  We also
note significant variations in the power-law flux over timescales similar to the
cold absorber, both within and
between the three observations.  The warm absorber does not vary
significantly within observations, but does show variations in column
density of a factor of $\geq10$ on timescales $\leq2$ weeks that seem
unrelated to the changes in the continuum, reflection or cold
absorber.  By assuming a uniform iron abundance for the reflection and
absorption, we have also established that ${\rm
  Fe/solar}=3.5^{+0.3}_{-0.1}$ is sufficient to model the broad-band
spectrum without invoking an additional partial-covering absorber.
Such a measurement is consistent with previous published constraints from the
2008 {\it Suzaku} observation alone, and with results from other
Seyfert AGN in the literature.
\end{abstract}


\section{Introduction}
\label{sec:intro}

In recent years, evidence has been mounting that the putative ``dusty
torus'' of active galactic nuclei (AGN)  unification schemes
\citep{Antonucci1993,Urry1995} is not a
homogeneous structure.  X-ray studies of several Seyfert AGN have
demonstrated significant variability in the column density and/or
covering fraction of the cold absorbing gas at typical radii of 
$r \geq 10,000\,r_{\rm g}$ from the supermassive black hole 
\citep{Risaliti2002,Risaliti2005a,Risaliti2010,Risaliti2011a}.
Physical interpretations for this inhomogeneity include a
patchy structure for the torus itself (e.g., \citealt{Turner2011}), a
system of comet-shaped, $\sim$neutral gas clouds orbiting the black hole at
Keplerian velocities within the Broad Emission Line Region (BELR; e.g.,
\citealt{Maiolino2010}), or perhaps a dust-driven wind beyond the dust
sublimation radius powered by radiation pressure from the disk
emission \citep{Czerny2012}.   
 
For black holes accreting at even a modest
rate ($L/L_{\rm Edd} \geq 0.01$, as per \citealt{Fender2010}), the
accretion flow is expected to remain
optically-thick and only moderately ionized down to at, or very near,
the innermost stable circular orbit (ISCO) in the disk
\citep{Reynolds2008}.  As such, if the standard picture of AGN
X-ray emission is correct (i.e., coronal Compton scattering of the thermal
disk photons, producing both the continuum power-law, and reflection
spectrum from this scattered power-law emission reprocessed by the
inner disk, e.g., \citealt{Reynolds2003,Miller2007}), we should expect
to see inner disk reflection features in any actively-accreting AGN in which our line of
sight to the inner disk is not completely obscured.\footnote{Note,
  however, that not all type 1 AGN that meet this requirement do show such features: X-ray
  surveys of hundreds of bright, nearby type 1 AGN indicate that only $\sim40\%$
  have broad Fe K$\alpha$ lines, as per
  \citet{Nandra2007} and \citet{dlcP2010}, while even detailed, multi-epoch
  examinations of single AGN show potential evidence for an ephemeral
  presence of broad Fe K$\alpha$ \citep{Brenneman2012}.}  The most
prominent example of these reflection features is the
relativistically broadened Fe K$\alpha$ line at $6.4 \keV$.  This
broad emission line was first detected in the Seyfert 1 AGN
MCG--6-30-15 with {\it ASCA}, as reported by \citet{Tanaka1995}, and
has been confirmed by several successive missions with higher spectral
resolution, e.g., {\it XMM-Newton} \citep{Fabian2002} and {\it Suzaku}
\citep{Miniutti2007,Chiang2011}.  

If we are
looking through a patchy/cloudy cold absorbing medium (even one that
is Compton-thick), we may view the broad Fe K$\alpha$
line from the innermost accretion disk only intermittently.  This
opens up a new window of opportunity
in the search for signatures of relativistic effects in AGN: an
obscuring cloud covers/uncovers different parts of the accretion disk
at different times, directly probing the nature of the
disk emission.  In particular, the combination of gravitational
redshift and relativistic Doppler boosting should imply strong
differences between the spectral appearance of the receding and
approaching parts of an inclined,
thin disk.  For an ``eclipse'' event of sufficient length, column
density and contrast, and
with enough signal-to-noise (S/N), it is possible to obtain time-resolved spectra of the
eclipse as different chords of the disk are covered/uncovered \citep{Risaliti2011b}.
Examining the variation of the continuum and broad Fe K$\alpha$
profile during such an event would yield definitive proof/disproof of the
origin of the Fe K$\alpha$ line as an inner disk reflection feature.
It would also provide important constraints on the size of the inner
disk/corona system and the physical properties and location of the
eclipsing cloud.  Finally, it would enable us to perform tomography of
the inner accretion disk for the first time, tracing its physical
properties as a function of radius.

In this paper, we report on three {\it Suzaku} observations of the Seyfert
1.8 galaxy NGC~1365 ($z=0.00547$\footnote{NED average redshift;
  http://ned.ipac.caltech.edu}).  NGC~1365 is the first
and most studied case of a patchy/cloudy cold absorbing medium
($N_{\rm H} \sim 10^{23-24} \pcmsq$)
occulting the inner accretion disk/corona system with a duty cycle of $\sim50\%$
\citep{Risaliti2002,Risaliti2005a,Risaliti2007,Risaliti2009a,Risaliti2009b,Risaliti2010,Maiolino2010}.
Mass estimates for the supermassive black hole range from $M_{\rm BH}
\sim 2 \times 10^6 \Msun$ to $1 \times 10^8 \Msun$
(\citealt{Risaliti2009b} and references therein), with a moderate
Eddington ratio which is correspondingly uncertain ($L/L_{\rm Edd}
\sim 0.02-0.12$, based on bolometric luminosities given in
\citealt{Vasudevan2010} and the masses quoted above).
In addition to its varying cold absorber and continuously-seen broad
Fe K$\alpha$ line, {\it XMM-Newton} observations have revealed the
presence of an ionized absorbing wind, manifesting as
blueshifted absorption lines of Fe\,{\sc xxv} and Fe\,{\sc xxvi} from
$6.7-8.3 \keV$.  This wind varies in 
outflow velocity over $\geq$weeks-to-months-long timescales
\citep{Risaliti2005b}.  Hosted in a barred spiral galaxy of type
SB(s)b \citep{DeV1991}, NGC~1365 also contains significant extended
X-ray emission from a circumnuclear starburst, as imaged by {\it
  Chandra} \citep{Wang2009}.  This
emission dominates the X-ray spectrum below $\sim2 \keV$, while the
spectrum above $\sim3 \keV$ is dominated by the AGN
itself.

The first {\it Suzaku} observation,
taken in 2008, is described in detail in \citet{Risaliti2009c},
\citet{Maiolino2010} and \citet{Walton2010}.  We re-examine its
spectral and timing properties in the context of our two 2010
observations in order to assess the variability of the various
spectral components of this AGN on timescales ranging from hours to
years.  Our goal is to determine whether eclipses of the inner
disk/corona by cold gas take place during our new observations, and to
use these events to probe the structure of the nucleus and the nature of
the inner accretion flow in NGC~1365.  We describe our observations
and data reduction in \S2, our timing analysis in \S3, and our
time-averaged and time-resolved spectral analysis in \S4.  A
discussion of our results and our conclusions are presented in \S5
and \S6, respectively.  Throughout the paper, we assume cosmological
parameters of $H_0=70 \kmpspMpc$, $q_0=0$ and $\Lambda_0=0.73$.

\section{Observations and Data Reduction}
\label{sec:obs}

{\it Suzaku} \citep{Mitsuda2007} has observed NGC~1365
quasi-continuously on three occasions: January 2008
(160 ks), June 2010 (150 ks; 2010a) and July 2010 (300 ks; 2010b).  The 2008
observation has been discussed at length by \citet{Risaliti2009c},
\citet{Maiolino2010} and \citet{Walton2010}, so its data reduction
will not be detailed here, other than to say that we followed the
reduction steps taken by Maiolino and Walton, but included the latest
calibration updates.  Here we focus on the reduction of the two 2010
observations, which were observed two weeks apart in a single
campaign to search for the putative eclipses of the inner accretion
disk that have been noted in previous observations of NGC~1365
\citep{Risaliti2005a,Risaliti2007,Risaliti2009a,Risaliti2009b,Maiolino2010}.   

2010a took place from June 27-30, while
2010b extended from July 15-22.  In both instances the telescope
was trained on NGC~1365 in the X-ray Imaging Spectrometer (XIS)
nominal pointing position.  The data from the
three operational XIS
instruments (XIS~0, XIS~1 and XIS~3; \citealt{Koyama2007}) were reprocessed using the {\tt
  xispi} script in accordance with the {\it Suzaku} ABC
Guide\footnote{http://heasarc.gsfc.nasa.gov/docs/suzaku/analysis/abc/}
along with the latest version of the CALDB (as of December 20, 2011).
This reprocessing eliminated Earth
occultations, South Atlantic Anomaly (SAA) passages, and other high
background periods.  Light curves and spectra from all three detectors
were then extracted using
{\sc xselect}, with circular source regions $\sim200$'' in radius
centered on the nucleus.  Background regions were made as large as
possible while avoiding contamination from both the AGN and the
calibration sources in the corners of each detector.

For the XIS spectra, we combined
data from the front-illuminated (FI) detectors XIS~0+3 data using the
{\tt addascaspec} script in order to increase S/N.  The
FI and back-illuminated (BI; XIS~1) XIS spectra, responses and backgrounds were then rebinned to $512$
spectral channels from the original $4096$ using {\tt rbnpha} and {\tt
  rbnrmf} in order to speed up
spectral model fitting without compromising the resolution of the
detectors.  Finally, the XIS spectra were grouped to a minimum of $25$
counts per bin using {\tt grppha} in order to facilitate robust $\chi^2$ fitting.
The merged, background-subtracted, time-averaged FI spectrum of
2010a (2010b) has a net $0.7-12 \keV$ 
count rate of $0.180 \pm 0.001 \, (0.134 \pm 0.001) \ctsps$ for a total of $59,913$
($92,736$) counts.  The total number of $2-10 \keV$ counts is $40,871$ ($55,285$).
The total BI count rate is $0.186 \pm 0.001 \, (0.148 \pm 0.001) \ctsps$ for a total of
$37,226$ ($62,326$), or $20,038$ ($28,783$) counts when restricting to $2-10
\keV$.  See Table~\ref{tab:obs} for details.  For all of the fitting
presented in this paper we allow for a 
global flux cross-normalization error between the FI and BI spectra.
The BI/FI cross-normalization is fixed at 1.03, in accordance with its expected
value\footnote{{\it Suzaku} Memo 2008-06:
  http://heasarc.gsfc.nasa.gov/docs/suzaku/analysis/watchout.html}.  

For the Hard X-ray Detector (HXD; \citealt{Takahashi2007}), the PIN
instrument detected NGC~1365 in both
observations, though the GSO did not.
Data from PIN were first reprocessed using the {\tt aepipeline} script, and
were then reduced as per the {\it Suzaku} ABC Guide.
For background subtraction, we used the ``tuned'' non X-ray background
(NXB) event files for June and July 2010 from the {\it Suzaku} CALDB, along with
the appropriate response files and flat field files for epoch 9 data.
The NXB background in 2010a (2010b) contributed a
count rate of $0.479 \pm 0.001 \, (0.497 \pm 0.001) \ctsps$ to the 
total X-ray
background from $14-40 \keV$, our energy range for adequate S/N.
We modeled the cosmic X-ray background (CXB) contribution as per the
ABC Guide, simulating its spectrum in {\sc xspec} \citep{Arnaud1996}.
The simulated CXB spectrum contributed a count rate of $0.0223 \pm
0.0001 \, (0.0221 \pm 0.0001) \ctsps$ to the total X-ray background from
$14-40 \keV$.  The NXB and
CXB files were combined to form a single PIN background spectrum.   
In comparison, the PIN data had a count rate of $0.540 \pm
0.002 \, (0.564 \pm 0.002) \ctsps$ over the same
energy range.  
 
Because the PIN data only contain 256 spectral channels (vs. the 4096
channels in the unbinned XIS data), rebinning to 25 counts per bin was
not necessary in
order to facilitate $\chi^2$ fitting.  Rather, we rebinned the PIN
spectrum to have a S/N of 5 in each energy bin,
which limited our energy
range to $14-40 \keV$.  After reduction, filtering and background
subtraction, the PIN spectrum of 2010a (2010b) had a
net $14-40 \keV$ count rate of $0.084
\pm 0.003 \, (0.066 \pm 0.003) \ctsps$, equating to a total of $12,000$ $(21,000)$ counts
over this energy range (see Table~\ref{tab:obs}).  We also added $3\%$ systematic errors to the PIN
data to account for the uncertainty in the non-X-ray background data
supplied by the {\it Suzaku} calibration team.
For the spectral fitting presented in this paper,
we assume a PIN/XIS-FI cross-normalization factor of 1.16 as per the 
{\it Suzaku} memo
2008-06\footnote{http://heasarc.gsfc.nasa.gov/docs/suzaku/analysis/watchout.html}.

\begin{table}
\begin{center}
\begin{tabular}{|l|l|l|l|l|l|}
\hline \hline
{\bf Date} & {\bf Time (ks)} & {\bf XIS cts/s} & {\bf XIS cts} & {\bf PIN cts/s} & {\bf PIN cts} \\
\hline \hline
21 Jan. 2008 & 160 & $0.33$ & $52,800$ & $0.10$ & $16,000$ \\
\hline
27 Jun. 2010 & 150 & $0.18$ & $59,913$ & $0.08$ & $12,000$ \\
\hline
15 Jul. 2010 & 300 & $0.13$ & $92,736$ & $0.07$ & $21,000$ \\
\hline \hline
\end{tabular}
\end{center}
\caption{\small{Observation data for the three {\it Suzaku} datasets of
  NGC~1365 discussed in this work.  Count rates and total counts are
  over the $0.7-12 \keV$ energy band for the XIS instrument, and
  represent the co-added XIS-FI data only.
   Count rates and total counts are taken over 
  the $14-40 \keV$ band for the PIN instrument.}}
\label{tab:obs}
\end{table}

\section{Timing Analysis}
\label{sec:timing}

\begin{figure}
\centerline{
\includegraphics[width=0.35\textwidth,angle=270]{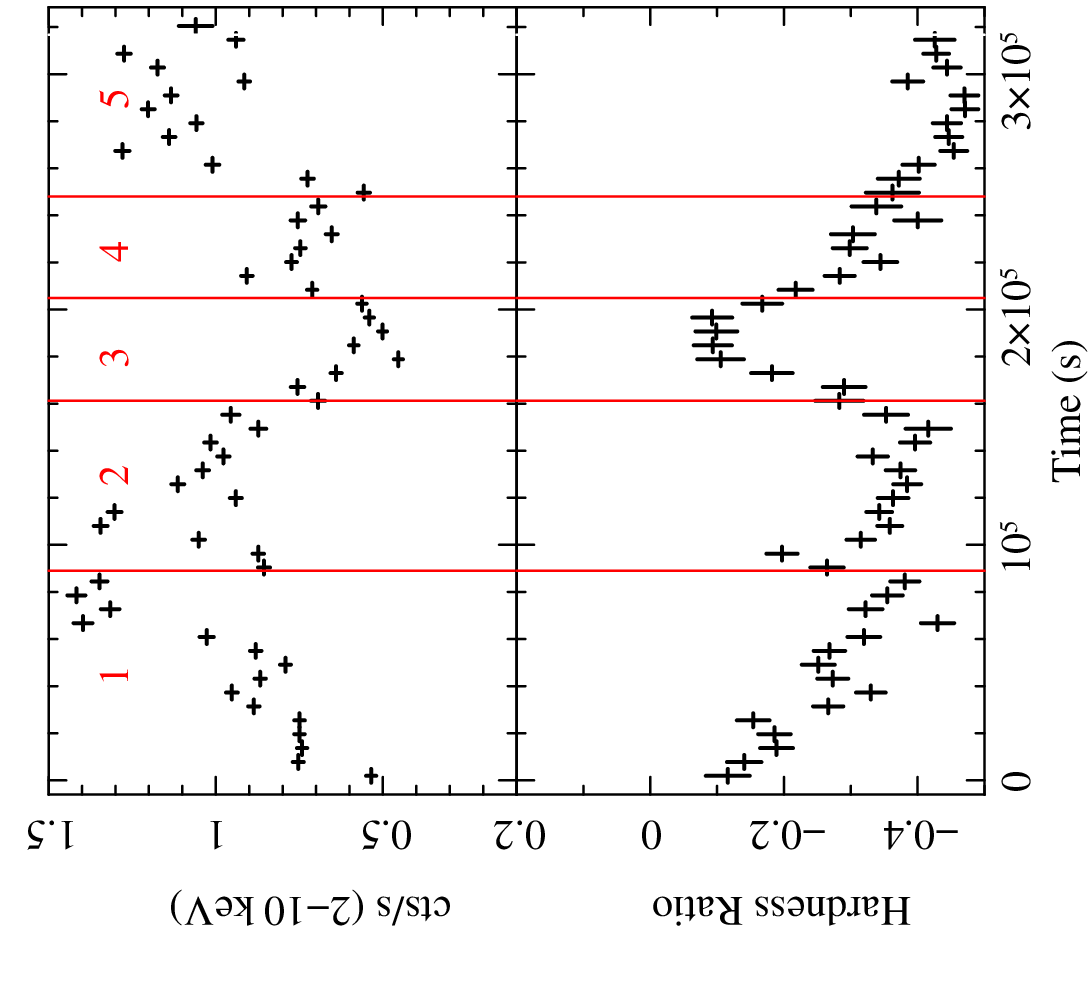}
}
\centerline{
\includegraphics[width=0.35\textwidth,angle=270]{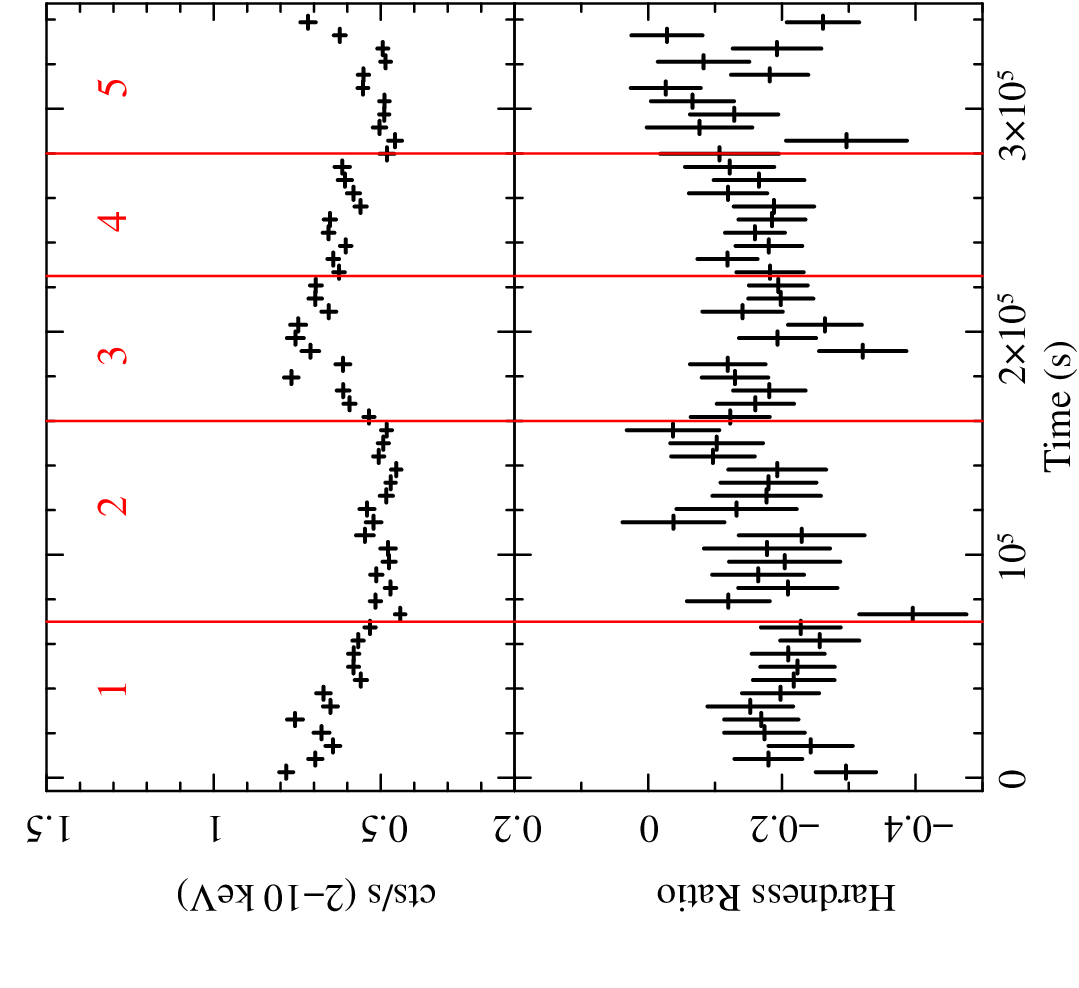}
}
\centerline{
\includegraphics[width=0.35\textwidth,angle=270]{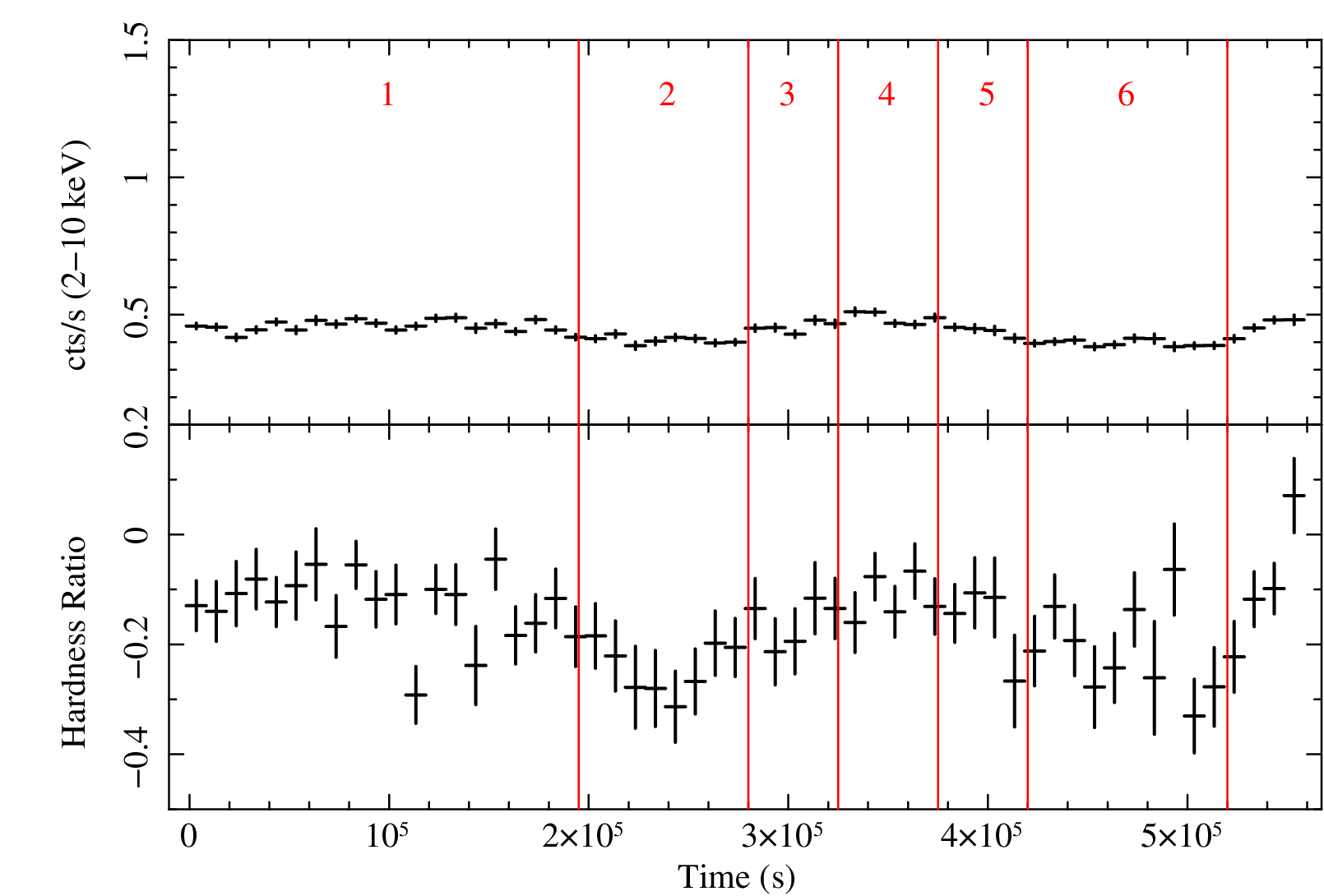}
}
\caption{{\small Light curves and hardness ratios 
    for the three {\it Suzaku} observations of NGC~1365.  Time bins
    are $5.9 \ks$ for 2008 and 2010a and $11.8 \ks$ for 2010b due to
    its longer exposure time.  Hardness
    ratio is defined as (hard-soft)/(hard+soft) flux, where the hard
    band extends from $6-10 \keV$ and the soft band covers $2-5
    \keV$.  {\it Top:}
    The 2008 observation is divided into five time intervals based
    primarily on source flux; these
intervals are shown by the red vertical lines and numbers.  Note the 
    anti-correlation between source flux and hardness ratio, a strong
    indicator of eclipse events.  {\it Middle:} 2010a
    also has five time intervals; intervals \#2 and \#5 are
periods of lowest flux.  Overall the source is less variable in flux
    than in 2008, and the hardness ratio is correspondingly less
    variable.  {\it Bottom:} 2010b is divided into
six time intervals based on flux, with periods of lowest flux in
    intervals \#2 and \#6.  The observation is roughly twice the
    length of 2008 and 2010a.  2010b has the lowest source flux of the
    three observations and also the least amount of variability in
    both flux and hardness ratio.}}
\label{fig:lc_hr}
\end{figure}

The light curves and hardness ratios for the three {\it Suzaku} observations of NGC~1365
are shown in Fig.~\ref{fig:lc_hr}.  The count rate of the 2008 data
varies by a factor of $\sim 3$, whereas the count rate of the 2010a
data varies by factor $\sim 1.8$ and the count rate of the 2010b data,
which has the lowest overall
flux of the three datasets, varies by only factor $\sim 1.3$.  This
pattern demonstrates, at least to first order, the expected
correlation between source flux and
variability \citep{Uttley2001} in actively accreting black hole systems.  

\begin{figure}
\centerline{
\includegraphics[width=0.35\textwidth,angle=270]{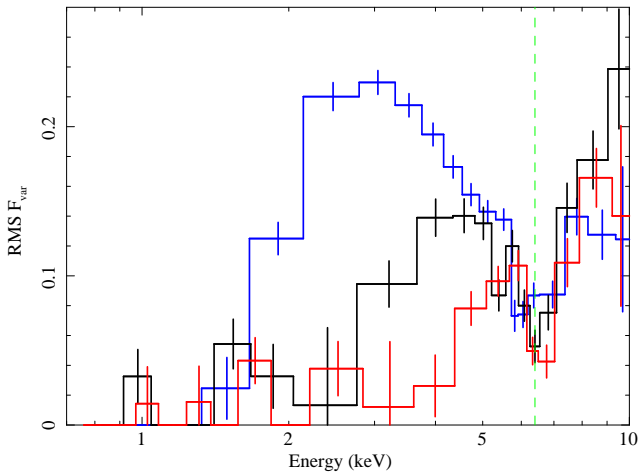}
}
\caption{{\small RMS $F_{\rm var}$ spectra for the combined XIS data from 2008
    (blue, highest), 2010a (black, middle) and 2010b (red, lowest).
    Note the lack of variability at energies below $\sim 2 \keV$, where the
    distant starburst emission dominates.  The increase in variability is
    noteworthy at energies above this point, which are dominated by a
    combination of the continuum, reflection and intrinsic absorption;
    the 2008 data are especially variable in this region.  The dip in variability
    at $\sim 6.4 \keV$ (marked by the dashed green line) seen in all
    three observations suggests that the Fe K$\alpha$ line has a substantial
    contribution from distant material that is slow to respond to continuum variations.
The overall increase in variability above $\sim 5 \keV$ in all observations
likely represents variable reflection from the inner accretion disk.  The
intrinsic absorber dominates the $3-5 \keV$ range, showing a steadily
    decreasing variability from 2008 through 2010a and 2010b}}
\label{fig:rms_fvar}
\end{figure}

It is evident that NGC~1365 is much
brighter and softer in 2008 than in 2010.  Note
also the clear anti-correlation
between source flux and hardness ratio in 2008 compared with the relatively
flat hardness ratios in 2010a and 2010b.  The presence of a cold
absorbing structure in the nucleus of NGC~1365 is well
established
\citep{Risaliti2005a,Risaliti2007,Risaliti2009a,Risaliti2009b,Risaliti2009c,Maiolino2010},
with variations in the cold column density by factors of up to $\sim
10$ having been reported between and within previous observations with {\it
  Chandra, XMM} and {\it Suzaku}.  
The observed trend in source flux and hardness ratio here is consistent
with this physical scenario.  In the framework of this
interpretation, we can infer that the source is relatively unobscured
in 2008, but shows distinct eclipses signified by spikes in the
hardness ratio coincident with dips in the X-ray light curve.  By contrast, the
source is both harder and dimmer
overall in 2010, decreasing in flux and increasing in hardness in
the two weeks between 2010a and 2010b, signifying an increasing column
density of the absorber.  Additionally, we note two other (though
more subtle than 2008) possible eclipse events in
each 2010 observation: intervals $\#2$ and $\#5$ in 2010a and $\#2$
and $\#6$ in 2010b.  Though these intervals mark times of lower source
flux, the hardness ratio during these intervals does not vary as
significantly as in the 2008 eclipses, implying either that the change in
column density during each 2010 event is not as great as that of the
eclipses on 2008, or that the column density itself is high enough that
its variation does not significantly impact the hardness ratio
(see. Fig.~\ref{fig:rms_fvar}).

The variation of the source flux vs. energy for the XIS data from each
observation is shown in Fig.~\ref{fig:rms_fvar}.  In each case,
the root-mean-square fractional variability (RMS $F_{\rm var}$) increases with
energy.  A high-energy
  component that varies more than the component best describing the
  low-energy emission strongly suggests the presence of inner disk
  reflection signatures in the spectrum.  The prominent dip in RMS $F_{\rm var}$
  that occurs between $6-7 \keV$ at
    approximately the location of the Fe K complex also indicates the
    presence of a narrow Fe
    K$\alpha$ line fluoresced from distant material (e.g., the outer
    disk or putative torus of Seyfert unification schemes), which
    would vary much less, and on timescales much longer than inner disk reflection
    signatures seen in and around the same energy range.  Most of the variation
    between the RMS $F_{\rm var}$ of the 2008 and 2010a observations takes place from $3-5
      \keV$, in the region dominated by the effects of the power-law continuum, 
      the cold absorbing column, and the broad red wing of the Fe K$\alpha$
      line.  The key question is which of these components is the primary driver of the
      spectral changes noted within and between
      these observations.  

\begin{figure}
\centerline{
\includegraphics[width=0.35\textwidth,angle=270]{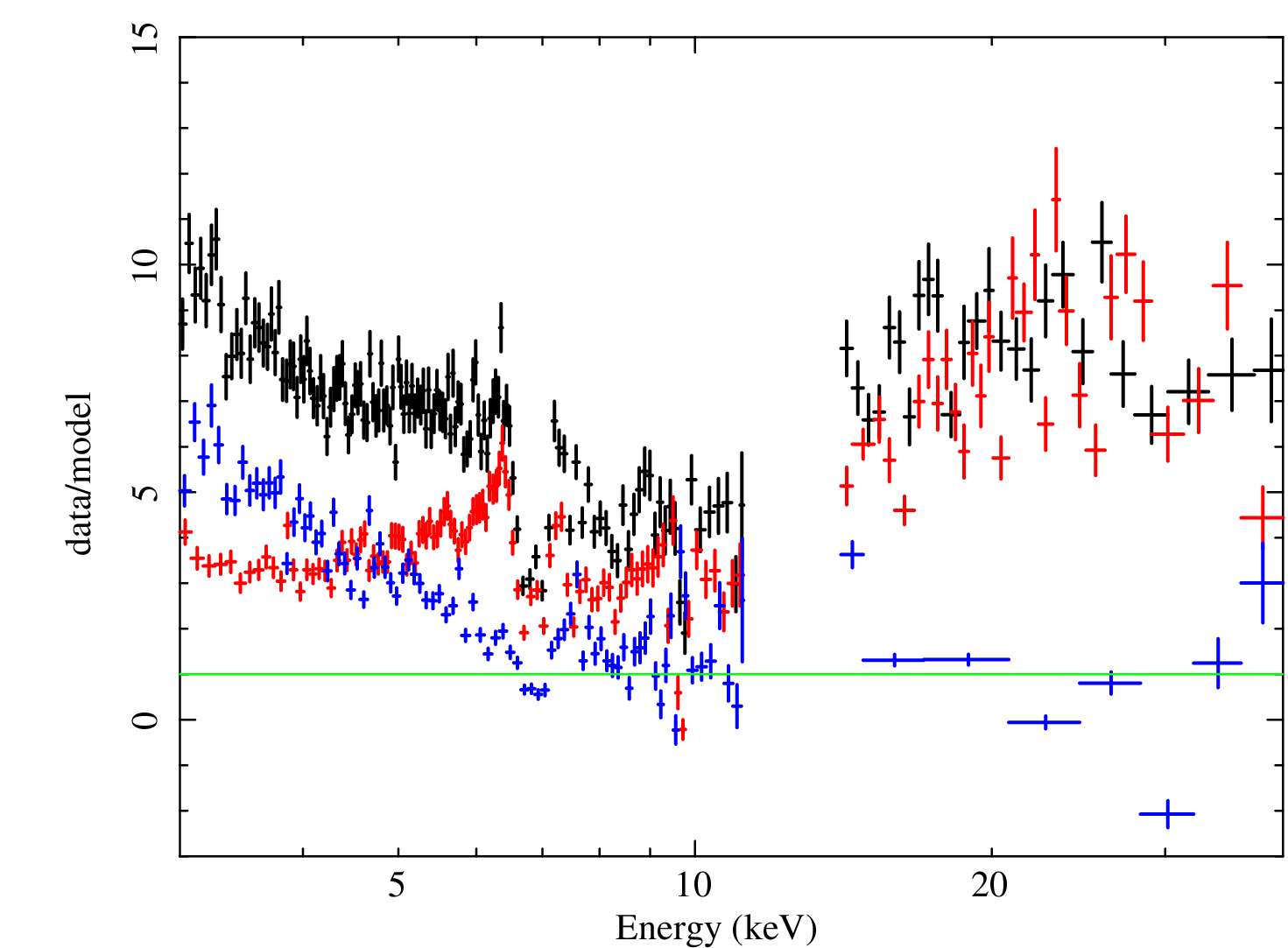}
}
\centerline{
\includegraphics[width=0.35\textwidth,angle=270]{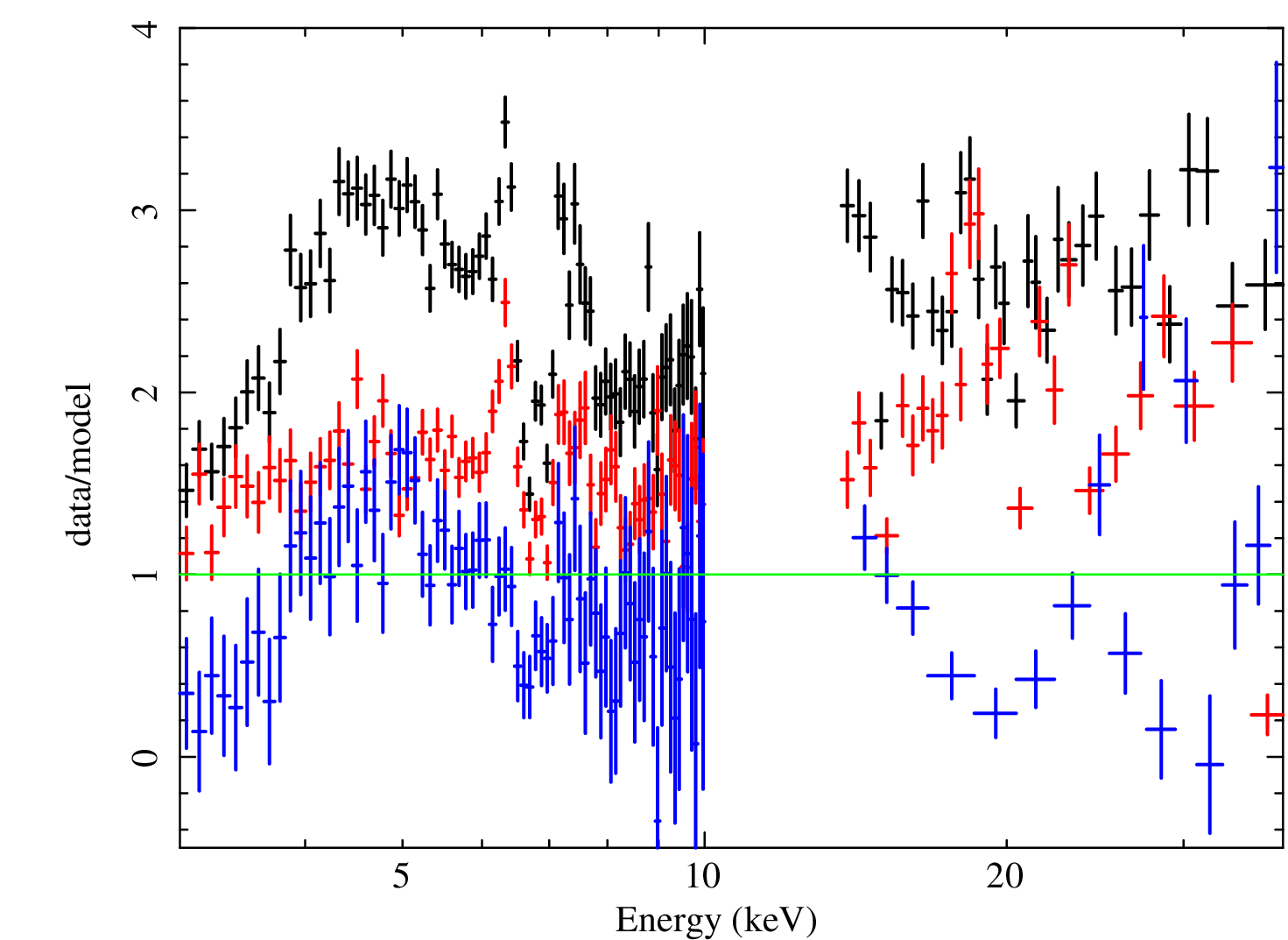}
}
\centerline{
\includegraphics[width=0.35\textwidth,angle=270]{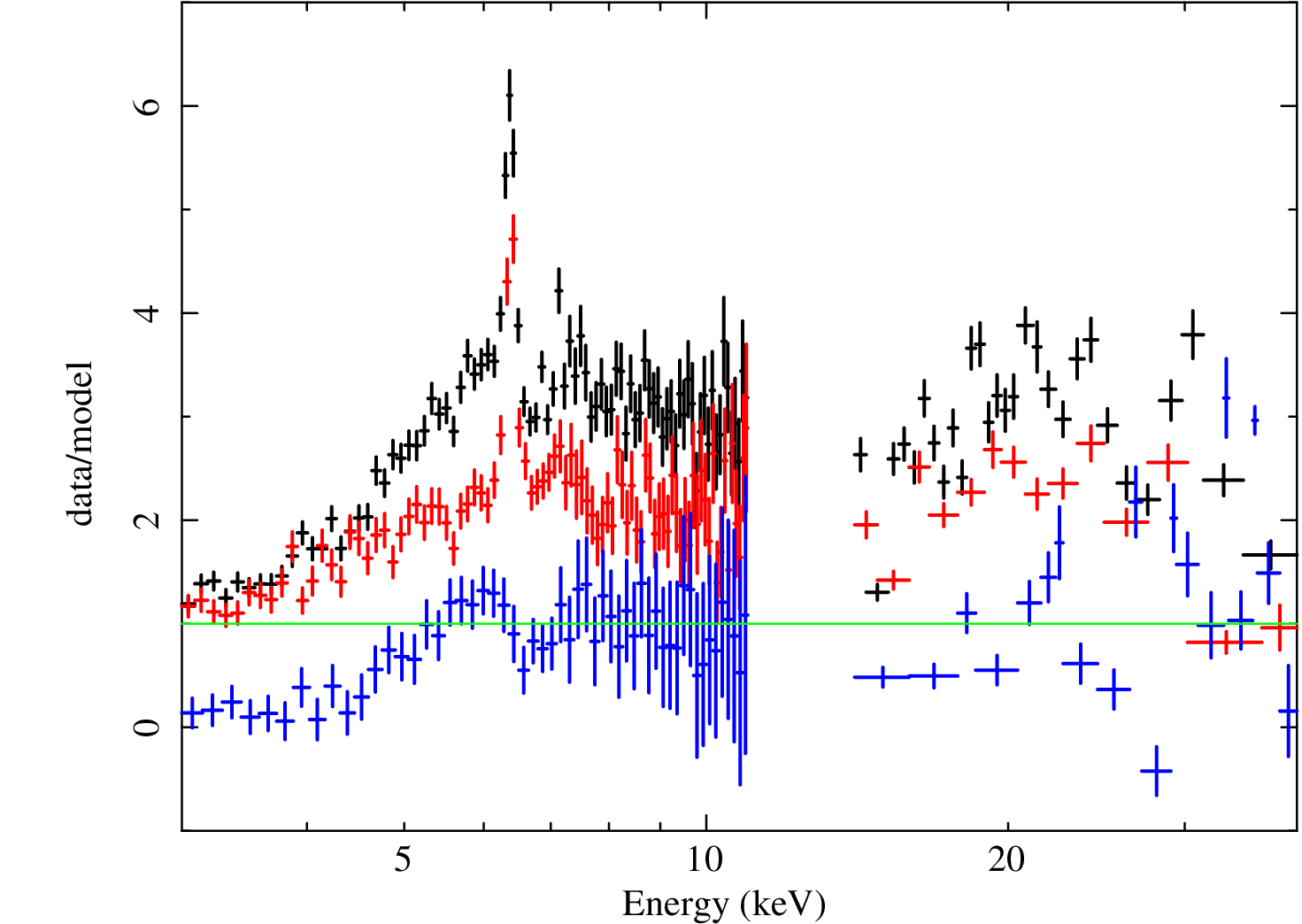}
}
\caption{{\small {\it Top:} High flux (black, highest) vs. low flux
    (red, middle) spectra from the
    five time intervals in the 2008 data,
    plotted along with the difference spectrum (high-low; blue,
    lowest).  The spectra are ratioed against an absorbed power-law,
    with the parameters fixed at their time-averaged valued except for
    the power-law normalization, which is allowed to vary freely.
    {\it Middle:} The same plot, now representing the 2010a data.  
{\it Bottom:} The same plot, now for the 2010b data.  
The change in spectral shape for the 2008 observation is well-modeled by a
change in the power-law flux and cold absorber column.
By contrast, the change in spectral shape for
    both 2010 observations is best
    accounted for not only by a change in the cold absorber column and power-law
    flux, but also by a
    modest change in the inner disk reflection.}}
\label{fig:diff_spec}
\end{figure}

Difference spectra for the three observations of NGC~1365 are shown in
Fig.~\ref{fig:diff_spec}, showing the variations in spectral shape
that result when the co-added, low-flux XIS and PIN spectra are subtracted from the
co-added high-flux XIS and PIN spectra.  
The change in spectral shape above $3 \keV$ for the 2008 observation is well-modeled by a
change in both the power-law flux and cold absorber column.
By contrast, the change in spectral shape for
    both 2010 observations is 
    accounted for almost entirely by a change in the cold absorber
    column, with only a minor contribution from a change in the inner disk reflection.
This conclusion will be
discussed in greater depth with our analysis of the spectral data in \S4.1.

\section{Spectral Analysis}

We begin our spectral analysis by examining the time-averaged spectra
of 2010a and 2010b in
conjunction with that of the 2008 data in order to examine the changes
in spectral shape and flux that the source exhibits over both
weeks- and years-long timescales.

The physical merit
of analyzing the time-averaged spectra is questionable because of
the eclipses by clouds of varying densities during the observations, and the changes in
spectral shape resulting from these occultations.  However, the high
S/N, time-averaged spectra can be used to
identify the various physical components that must be included in
order to model the data accurately.  We interpret the parameter values returned by
these fits as fiducial in nature, providing a starting point within
the parameter space for our subsequent time-resolved spectral
analysis, which provides a more accurate representation of the
variation of the physical parameters describing the nucleus of
NGC~1365. 

We explored the parameter space of each model we employed using the
Markov Chain Monte Carlo (MCMC) algorithm within {\sc xspec}.  This
procedure is especially effective for thoroughly probing complex,
multi-dimensional parameter spaces (e.g., \citealt{Reynolds2012}).  To assess the
parameter space of a given model, four independent, 55,000-element
chains were created, each beginning at a different random seed within
the parameter space.  The first 5000 elements of each chain were
discarded during a ``burn-in'' phase.  A diagonal, Gaussian proposal
was used to create the chains, beginning with the squares of the
$1\sigma$ {\sc xspec}-derived errors on the initial fit of the model
to the data.  The covariance matrix was then rescaled iteratively
until a parameter variance within the chain of $\sim0.75$ was achieved for
each run.  The Rubin-Gelman convergence criterion of $\sim1$ was
achieved for each chain.  All errors quoted for our resulting spectral
fits are derived from the combined $200,000$-element chain and
correspond to $90\%$ probability, unless otherwise specified.

\subsection{Time-averaged Spectra}
\label{sec:TA}

\begin{figure}
\centerline{
\includegraphics[width=0.35\textwidth,angle=270]{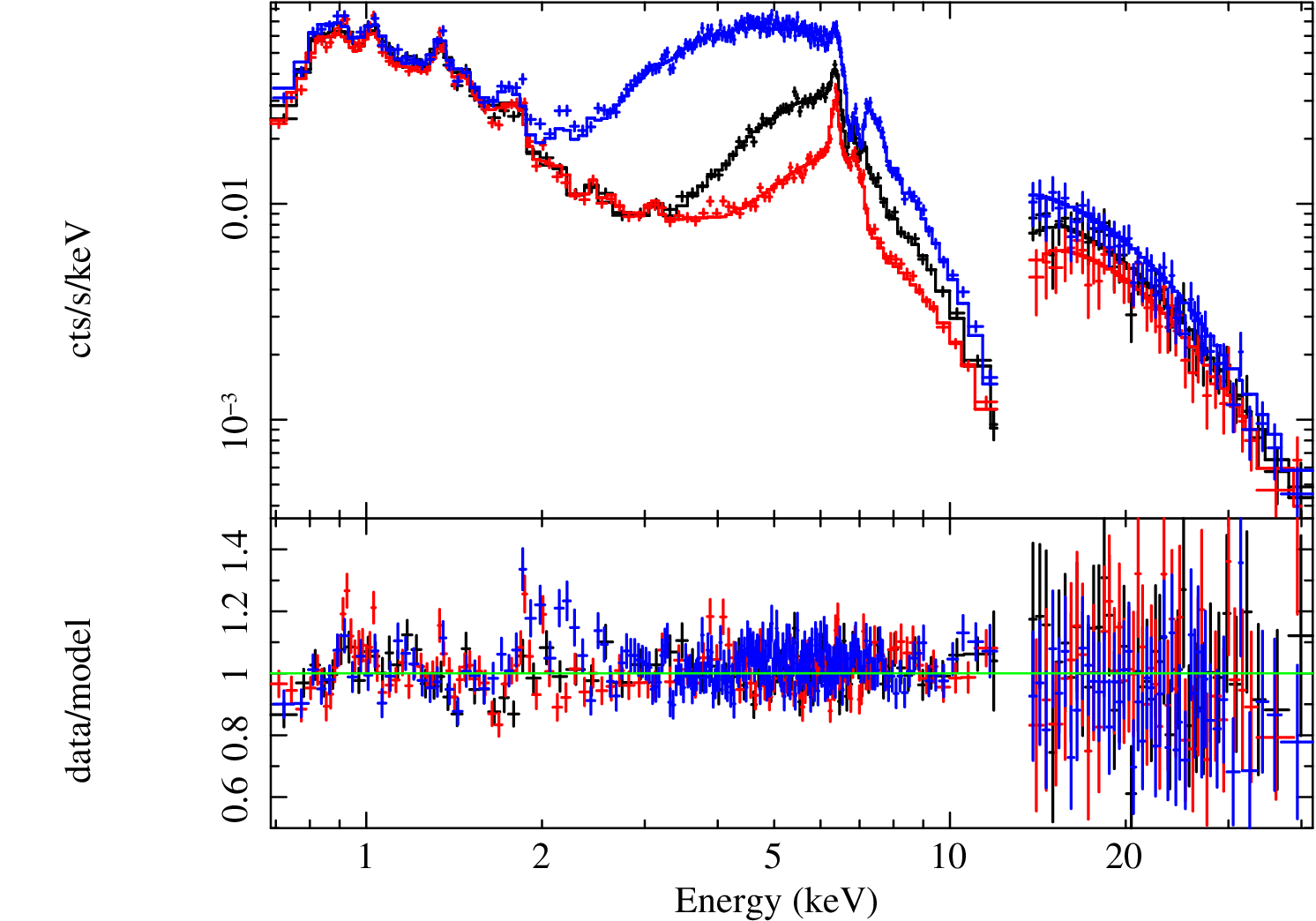}
}
\vspace{0.5cm}
\centerline{
\includegraphics[width=0.35\textwidth,angle=270,bb=80 -5 581 700,clip=]{fig4b.eps}
}
\caption{{\small {\it Top:} Time-averaged spectra for the data from 2008,
    (XIS-FI and PIN in blue, highest), 2010a (XIS-FI and PIN in black,
    middle) and
    2010b (XIS-FI and PIN in red, lowest).  The
    data/model ratio with respect to our best-fitting model
    (Table~\ref{tab:TA_tab}) is shown in the lower panel (green line
    indicates a theoretical perfect fit).  These residuals are dominated by calibration
    features in the $1.5-2.5 \keV$ range, especially, which we ignore in our model
    fitting but show here for completeness.  {\it Bottom:}
    Data-to-model ratio plot for the Fe K region, this time with the
    WA, broad Fe K$\alpha$ and narrow Fe emission and absorption lines
    removed from the best-fit model in order to show the residuals of
    these features.  Color scheme is the same as for the top panel, with
    emission and absorption lines labeled in green.  The lines connecting data
    points are meant to guide the eye and do not represent a model.
    Note the more prominent broad Fe K$\alpha$ line in 2008 and the
    more prominents narrow Fe K$\alpha$ line in 2010b.}}
\label{fig:TA_spec_data}
\end{figure}

\begin{figure}
\centerline{
\includegraphics[width=0.35\textwidth,angle=270]{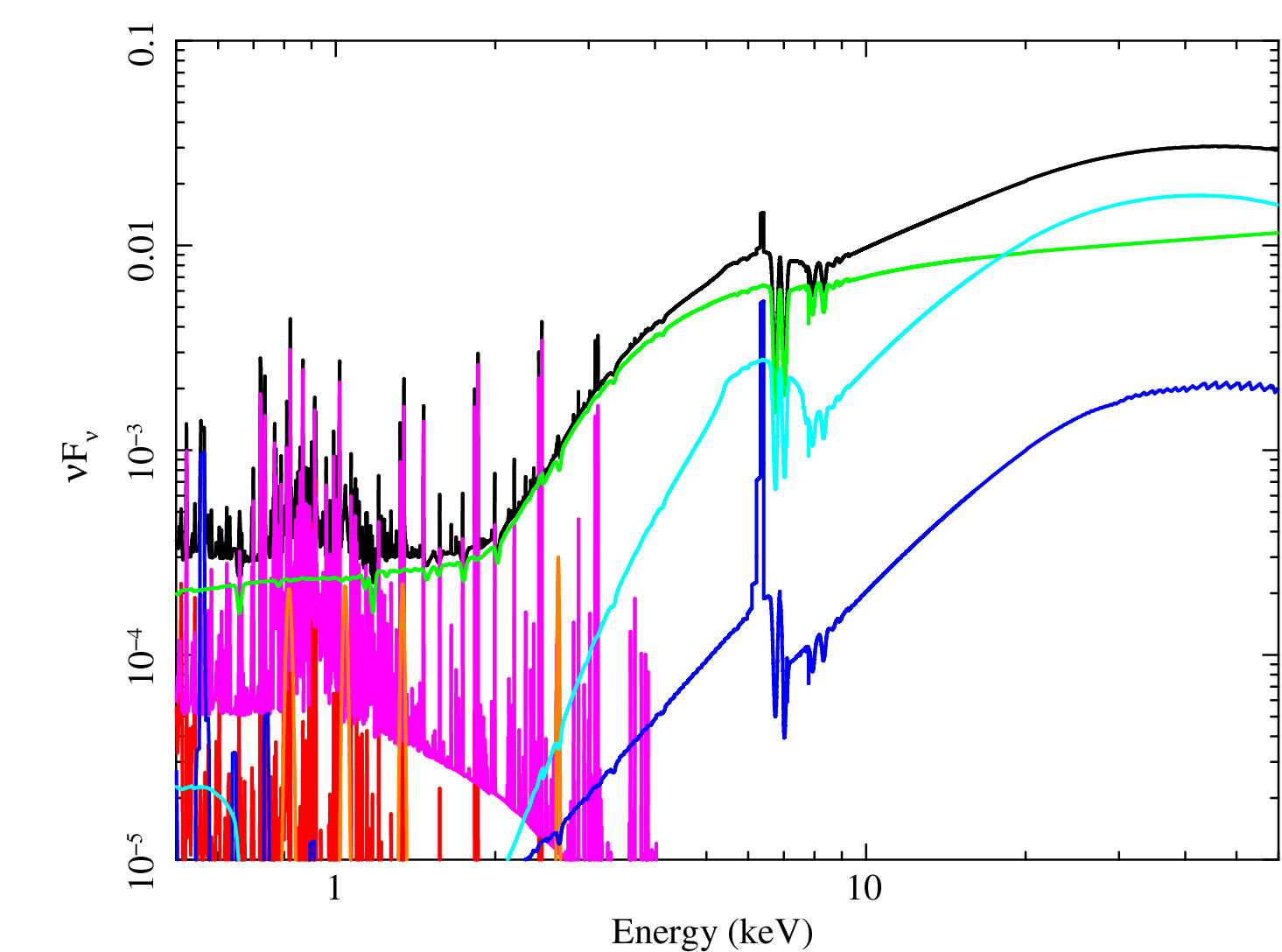}
}
\centerline{
\includegraphics[width=0.35\textwidth,angle=270]{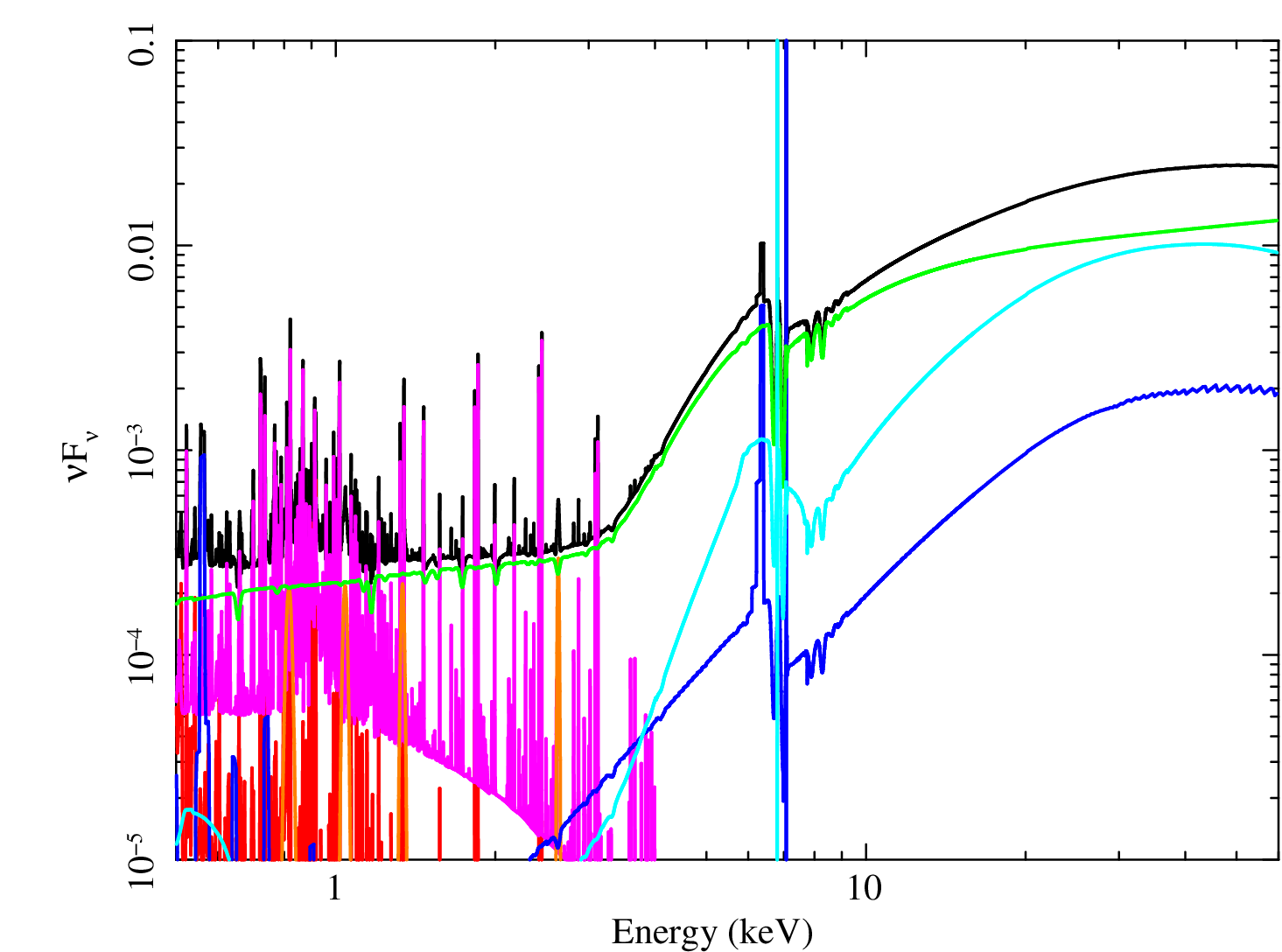}
}
\centerline{
\includegraphics[width=0.35\textwidth,angle=270]{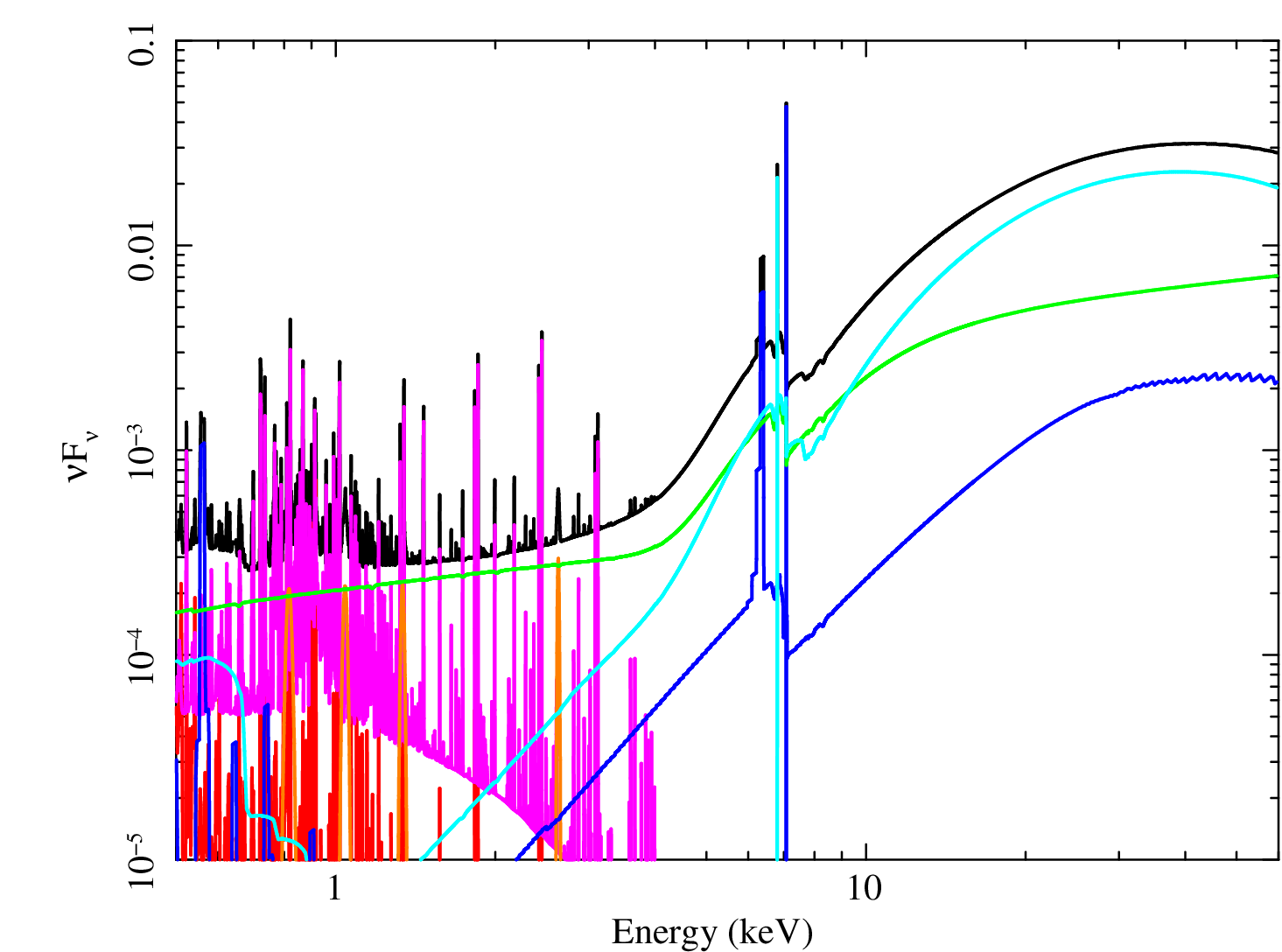}
}
\caption{\small{{\it Top:}
    $\nu F_{\nu}$ plot of the best-fitting model components for the
    2008 data.  The absorbed power-law making up most of the model flux
    is in green, the two-temperature thermal starburst plasma is shown
    below $\sim3 \keV$ in red and
    magenta, distant reflection is in dark blue and inner disk reflection is in
    light blue.  The total spectral model is in black.  {\it Middle:} Same
    as left panel, but for the
    2010a data.  {\it Bottom:} Same as left panel, but for the
    2010b data.  Note that the starburst emission is constant between
    observations, with variability only seen in the spectrum above $3 \keV$.}}
\label{fig:TA_spec_eemo}
\end{figure}

An examination of the time-averaged spectra extracted from each of the three
{\it Suzaku} observations of NGC~1365 presents several features which are
readily apparent, over and above the standard power-law continuum describing all
AGN: a constant soft component below $2 \keV$ due to a
circumnuclear starburst component (resolved by {\it Chandra} and detailed in \citealt{Wang2009}, but
not focused on here in favor of examining the properties of the
nucleus), a highly changeable region from
$\sim2-5 \keV$ in which the curvature is due to a variable cold absorbing column
of gas, narrow Fe K emission line(s) from $6-7 \keV$ with a subtle curvature on
the red wing suggesting the presence of a broad Fe K$\alpha$ line, variable narrow Fe K
absorption lines in the $6-7 \keV$ range indicative of ionized gas within the
system, and slightly variable curvature above $10 \keV$ due to reflection of the
primary continuum from the disk and/or distant torus.  These features can be
seen in the spectra and models shown in Figs.~\ref{fig:TA_spec_data}-\ref{fig:TA_spec_eemo}, and
the best-fit model and parameter values describing each observation
are shown in Table~\ref{tab:TA_tab}. 

Operationally, we model these physical components with the following functional form: 
{\tt TBabs*(2vapec+4zgauss+WA*(reflionx+\newline zpcfabs*(powerlaw+zgauss+relconv(reflionx))))}.\newline
This type of model and its components are commonly used to describe
AGN spectra showing similar reflection and absorption features to NGC~1365, e.g., MCG--6-30-15
\citep{Fabian2002,Brenneman2006,Chiang2011} and NGC~3783
\citep{Brenneman2011,Patrick2011,Reynolds2012}. 
The parameter values and $90\%$ confidence errors for each model component are presented in
Table~\ref{tab:TA_tab}, which represents a joint fit to all three time-averaged spectra.  
The two-temperature thermal plasma ($kT_1= 0.3 \keV,\, kT_2= 0.69 \keV$) and four
Gaussian lines ($E_1= 0.82 \keV, \, EW_1= 10 \eV, \, E_2= 1.05 \keV, \, EW_2= 12
\eV, \, E_3= 1.34 \keV, \, EW_3= 16 \eV, \, E_4= 2.64 \keV, \, EW_4= 24 \eV$) representing the
starburst emission are held constant at the parameter values determined by
\citet{Wang2009}.  

The warm absorber (WA) in NGC~1365 manifests primarily in the Fe K
band as a series of four prominent absorption lines (the K$\alpha$ and K$\beta$
lines of Fe\,{\sc xxv} and Fe\,{\sc xxvi}) which have been discussed
extensively in \citet{Risaliti2005b} and \citet{Risaliti2007}.  We model this
component with an {\sc xstar} (v2.2.0)\footnote{The latest release of
  the {\sc xstar} software (v2.2.1) has several updates to the atomic
  data for iron, which result in larger measured column densities when
using v2.2.1.  Using an
identical table model generated with v2.2.1, we note that the column
densities derived from the data increase by a factor of $\sim2$.} multiplicative table within
{\sc xspec}, generated specifically for NGC~1365 and identical to the
one used in \citet{Walton2010}.  The
free parameters are the column density and ionization parameter of the
gas, its redshift (a proxy for outflow velocity) and its iron
abundance (which we tie to the iron abundance of the reflection components).
This WA has a relatively high ionization of $\xi \geq
3000 \ergcmps$ in all three observations, though its column density increases by
a factor of $\sim2$ between 2008 and 2010a, then decreases by a factor of
$\sim10$ in the two weeks between 2010a and 2010b.  This variability appears to
be unrelated to the variability of other components in our time-averaged
spectral fitting.  Additionally, the absorption lines which comprise
the WA are all
blueshifted, as noted by \citet{Risaliti2005b,Risaliti2007}.  The outflow velocity of
the gas also varies between observations, going from $v_{\rm out} = 3651^{+231}_{-360}
\kmps$ in 2008 to $v_{\rm out} = 1884^{+285}_{-600} \kmps$ in 2010a
to $v_{\rm out} = 2961^{+560}_{-1380}
\kmps$ in 2010b.  Though the 2010 observations do not demonstrate significant
variability in two weeks, taking errors into account, the change in outflow velocity does
appear to be robust at the $\sim4\sigma$ level on two-year timescales.

The distant reflector is
modeled by a {\tt reflionx} component
\citep{Ross2005} with its ionization fixed at $\sim$neutral ($\xi=1 \ergcmps$), its
incident power-law index tied to that of the power-law component and its iron
abundance tied to that of the warm absorber.  The normalization of this
component does not vary significantly between the three observations.  In
addition to the narrow Fe K$\alpha$ and K$\beta$ emission lines that are the most prominent features of
this distant reflection, we also note a possible narrow line of ionized iron at $6.84
\keV$ in both of the spectra from 2010, though this feature is only
statistically significant in 2010a.  The line is unresolved and its width is
therefore fixed at $10 \eV$, but, if present, it would suggest that the distant reflector
is not entirely composed of neutral gas.  Allowing the ionization parameter of
the distant {\tt reflionx} component to vary yields only an upper limit of $\xi
\leq 1.2$, however, so we have elected to continue with $\xi$ fixed at unity. 

We note that the cold absorbing column ({\tt zpcfabs}) is quite
prominent in NGC~1365, averaging a density of $(1.5^{+0.3}_{-0.2}) \times
10^{23} \pcmsq$ in 2008 and increasing steadily to $(5.8^{+0.5}_{-0.6}) \times 10^{23} \pcmsq$
two years later in 2010a and $(10.7^{+0.5}_{-0.5}) \times 10^{23} \pcmsq$ two weeks later in
2010b.  It has a covering factor (denoted $f_{\rm cov}$) over the hard X-ray source and
inner disk, which is $(95 \pm 1)\%$ in 2008 and 2010a, but decreases
to $(91 \pm 1)\%$
in 2010b at the $4\sigma$ level.  Given its high value, we interpret this covering factor physically as a
scattered fraction, rather than as a formal partial-covering absorber. 

The photon indices of the power-law components do not vary
significantly ($\leq 2\sigma$)
between observations ($\Gamma \sim 1.75$), and the normalization of
the power-law is constant within $10\%$ from 2008 to 2010a, but drops
by a factor of $\sim2$ ($\geq 6\sigma$) in the two weeks between 2010a and 2010b.  

\begin{figure}
\centerline{
\includegraphics[width=0.35\textwidth,angle=270]{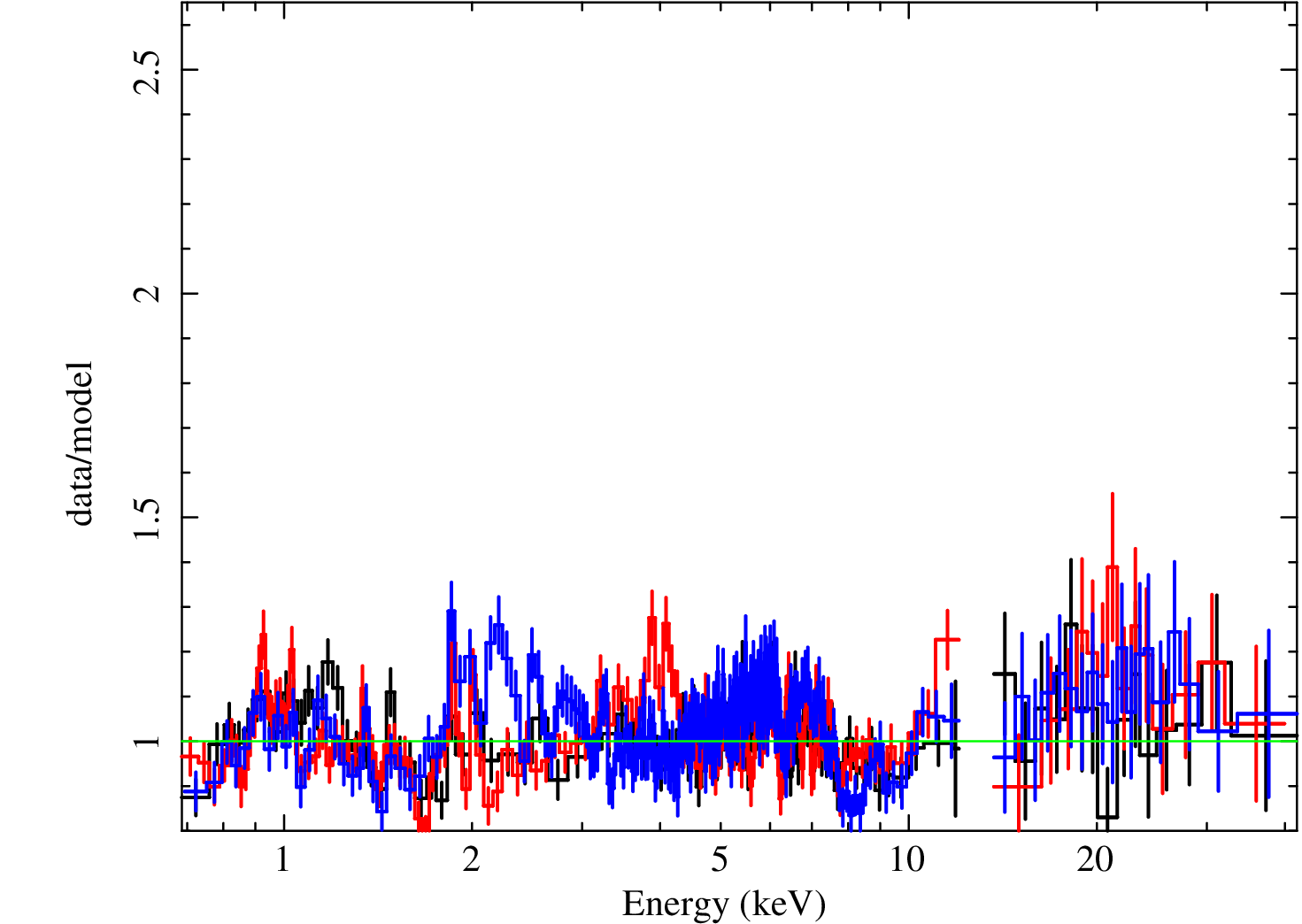}
}
\vspace{0.5cm}
\centerline{
\includegraphics[width=0.35\textwidth,angle=270,bb=80 -5 581 700,clip=]{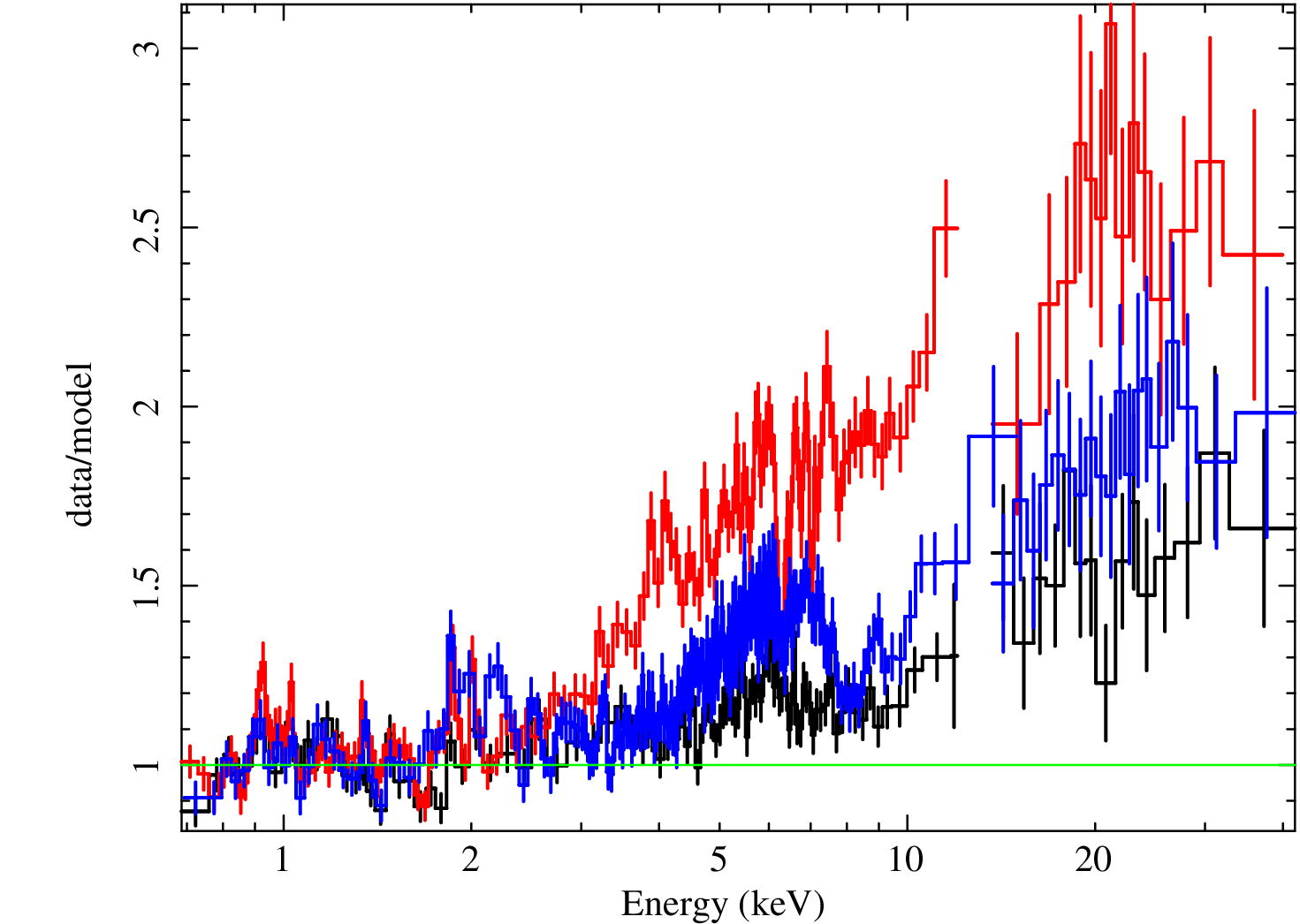}
}
\caption{{\small {\it Top:} The ratio for the best-fitting model to the data
    from the three {\it Suzaku} observations of NGC~1365, with the inner disk
    reflection component subtracted in order to illustrate the spectral features
    fitted by this component.  2008 data are shown in blue (middle),
    2010a in black (lowest) and 2010b data in red (highest).  {\it
    Bottom:} Same as the top plot, but this time
the best-fitting model without inner disk reflection has been
    refitted, yielding a significantly worse fit by $\Delta\chi^2/\Delta\nu=+385/+14$ compared with
    the best fiting model.
This demonstrates the importance of this component to the global
    goodness-of-fit.}}
\label{fig:best_nordc_ratio}
\end{figure}

The inclusion of an inner disk reflection component, parametrized
above by {\tt relconv(reflionx)}, is necessary in order to achieve a
statistically adequate model fit to the data in each of our three
time-averaged observations, in spite of the variation in spectral
state between them.  Eliminating this component and refitting the
time-averaged data worsens the global
goodness-of-fit by $\Delta \chi^2/\Delta \nu=+385/+14$ and clearly
results in a poor fit to the eye with large residuals in the $3-10
\keV$ region (see Fig.~\ref{fig:best_nordc_ratio}).

As with the distant reflector, we have tied the iron abundance of the inner disk
reflector to that of the warm absorber, and the index of the incident power-law
radiation to that of the power-law component.  The ionization of the
inner disk reflector does not appear to deviate
significantly from neutrality in any of the observations.  We further note that the
normalization of this component is nearly ten times that of the
distant reflector, but it does not vary significantly, dropping by a factor of $\sim1.8$
from 2008 to 2010a (at only $0.8\sigma$) and by an additional factor of $\sim1.2$ between 2010a and
2010b (at only $0.9\sigma$).  

We convolve the inner disk reflector with a {\tt relconv} smearing
kernel \citep{Dauser2010}, which imprints the effects of relativistic Doppler shifts
from the rotating disk and General Relativity from the central supermassive black hole onto the spectrum.
We assume that the disk is inclined at $i=58\degmark$, in keeping with the
results of \citet{Risaliti2005a,Risaliti2007,Risaliti2009c}, \citet{Maiolino2010} and
\citet{Walton2010}.  We also assume that the disk radiates $\propto r^{-q}$,
where $q$ is allowed to vary freely and the emitting portion of the disk extends
from $r_{\rm in}$ to $r_{\rm out}=400\,r_{\rm g}$ for a
maximally-spinning black hole, where $r_{\rm in}$ is also allowed to vary freely.  This
approach allows us to directly compare our results to previous work:
$q=5.0^{+1.0}_{-0.5}$ in the 2008 {\it Suzaku} observation
analyzed by \citet{Walton2010}, and it is measured between $q=5$ and
$q=9$ ---constant within errors--- in the three {\it Suzaku} observations
considered here (our measured $q=5.9^{+0.2}_{-0.3}$ in 2008 is entirely consistent with
Walton \etal, within errors).  Our measured inner disk radius, assumed
to be the ISCO, is constant within errors between 2008 and 2010a
($r_{\rm in}=1.9^{+0.1}_{-0.1}-2.0^{+0.1}_{-0.1}\,r_{\rm g}$), but drops to
$r_{\rm in}=1.3^{+0.2}_{-0.1}\,r_{\rm g}$ in 2010b.  This change is
only significant at the $1.8\sigma$ level, however.  We note that our 2008
measurement is consistent with that of Walton \etal as well ($r_{\rm
  in}=1.85 \pm 0.15\,r_{\rm g}$).  

The low inner disk radii measured from the data suggest a
rapidly-rotating, prograde black hole with a spin of $a \geq 0.93$, provided
that the inner edge of the disk measured here is indeed the ISCO.
If instead we
fix the inner radius of the disk to $r_{\rm in}=r_{\rm ISCO}$ for each observation
and allow the black hole spin to vary freely (linking the spin value between
observations), we measure $a=0.96 \pm 0.01$ (where $a \equiv cJ/GM^2$ for a
supermassive black hole of mass $M$ and angular momentum $J$).
Because the measurement of black hole spin in NGC~1365 is not the
focus of this paper, however, we do not perform a rigorous analysis of
the spin determination and sources of error here.  This analysis will
be presented in a forthcoming paper (Brenneman \etal 2013, in preparation).

An examination of the unabsorbed fluxes and luminosities of NGC~1365
in Table~\ref{tab:TA_tab} indicates that the source was brightest in 2008
and dimmest in 2010b, though the difference in unabsorbed power
(factor 1.4 between the three observations) is
less than that of the absorbed power (factor 2.0 between the observations), indicating that absorption
plays a significant role in shaping the spectrum of the AGN.  Indeed,
the change of the power-law normalization and inner disk reflection components alone
cannot account for the change in spectral shape between $\sim2-5 \keV$
between 2008, 2010a and 2010b.  Attempting
to account for the change entirely with the power-law and inner disk reflector results in a
worsening of the fit by $\Delta \chi^2/\Delta \nu=+3237/+2$ and a clearly
inadequate visual match to the data.  This reinforces the hypothesis
that the primary driver of the spectral shape change between the three observations
is the cold absorbing column.

Previous studies (e.g., \citealt{Risaliti2009c}) have cited the need
for a second cold absorber to explain the hard energy excess $\geq 10
\keV$.  We note that, as per \citet{Walton2010}, the requirement
for such a component
($f_{\rm cov}=100\%$) disappears if we allow the iron abundance of the
accretion disk to vary freely in our model fitting:
Fe/solar$=3.5^{+0.3}_{-0.1}$.  This value is driven jointly by the
presence of the Fe\,{\sc xxv} and Fe\,{\sc xxvi} K$\alpha$ and
K$\beta$ absorption lines, the strength and shape of the distant
reflection features and the strength and shape of the inner disk
reflection features (i.e., Fe K$\alpha$ line, Compton hump).  The {\sc
  xstar} table we used to model the WA and the {\tt reflionx} model
both allow the iron abundance to vary freely, and we tied the Fe/solar
values of all three components together in our best fit model, which
is confirmed by our rigorous MCMC analysis.
 
\onecolumn
\begin{table}
\begin{center}
\begin{tabular}{|ll|c|c|c|}
\hline\hline
{\bf Component} & {\bf Parameter (units)} & {\bf 2008} & {\bf 2010a} & {\bf 2010b}\\
\hline\hline
TBabs & $N_{\rm H}\,(\pcmsq)$ & $1.34 \times 10^{20}(f)$ & $1.34 \times 10^{20}*$
& $1.34 \times 10^{20}*$ \\
\hline
WA & $N_{\rm H}\,(\times 10^{22}\pcmsq)$ & $5.95^{+0.37}_{-0.34}$ &
$10.33^{+0.72}_{-1.32}$ & $\leq1.48$ \\
   & $\xi\,(\ergcmps)$ & $3019^{+369}_{-135}$ & $5495^{+814}_{-483}$ & $3019^{+696}_{-389}$ \\
   & $v_{\rm out}\,(\kmps)$ & $3651^{+231}_{-360}$ & $1884^{+285}_{-600}$ &
$2961^{+570}_{-1380}$ \\
\hline
Reflionx (distant) & Fe/solar & $3.5^{+0.3}_{-0.1}$ & $3.5*$ & $3.5*$ \\
         & $K_{\rm Rdist}\,(\phpcmsqps)$ & $3.9^{+0.3}_{-0.7} \times 10^{-5}$ &
$3.8^{+0.4}_{-1.0} \times 10^{-5}$ & $4.3^{+0.1}_{-0.2} \times 10^{-5}$ \\
\hline
Zpcfabs & $N_{\rm H}\,(\times 10^{22} \pcmsq)$ & $15.4^{+0.3}_{-0.2}$ & $58.3^{+0.5}_{-0.6}$ & $106.6^{+5.3}_{-3.2}$ \\
        & $f_{\rm cov}$ & $0.95^{+0.01}_{-0.01}$ & $0.95^{+0.01}_{-0.01}$ &
$0.91^{+0.01}_{-0.01}$ \\
\hline
Power-law & $\Gamma$ & $1.81^{+0.04}_{-0.04}$ & $1.76^{+0.02}_{-0.03}$ & $1.73^{+0.01}_{-0.13}$ \\
          & $K_{\rm PL}\,(\phpcmsqps)$ & $5.43^{+0.37}_{-0.27} \times 10^{-3}$ &
$4.92^{+0.20}_{-0.18} \times 10^{-3}$ & $2.30^{+0.41}_{-0.02} \times 10^{-3}$ \\
\hline
Zgauss & $E\,(\keV)$ & $6.84^{+0.13}_{-0.14}$ & $6.84*$ & $6.84*$ \\
       & $\sigma\,(\keV)$ & $0.01({\rm f})$ & $0.01*$ & $0.01*$ \\
       & $K_{\rm line}\,(\phpcmsqps)$ & $0.0^{+0.7}_{-0.0} \times 10^{-5}$ &
$2.3^{+0.7}_{-0.3} \times 10^{-5}$ & $0.5^{+0.2}_{-0.1} \times 10^{-5}$ \\
       & $EW\,(\eV)$ & $0^{+1}_{-0}$ & $93^{+28}_{-12}$ & $28^{+11}_{-6}$ \\
\hline
Relconv & $q$ & $5.9^{+0.2}_{-0.3}$ & $6.9^{+2.1}_{-1.0}$ & $5.4^{+0.3}_{-0.4}$ \\
        & $r_{\rm in}\,(r_{\rm g})$ & $1.9^{+0.1}_{-0.1}$ & $2.0^{+0.1}_{-0.1}$ &
$1.3^{+0.2}_{-0.1}$ \\
\hline
Reflionx (inner) & $\xi\,(\ergcmps)$ & $\leq 2.0$ & $\leq 2.0$ & $1.7^{+1.2}_{-0.4}$ \\
         & $K_{\rm Rrel}\,(\phpcmsqps)$ & $4.06^{+0.05}_{-2.10} \times 10^{-4}$
    & $2.34^{+0.04}_{-1.30} \times 10^{-4}$ & $3.50^{+0.10}_{-1.80}
\times 10^{-4}$ \\
\hline
$F_{\rm 2-10}$ & absorbed ($\ergpcmsqps$) & $1.27 \times 10^{-11}$ &
$6.13 \times 10^{-12}$ & $4.05 \times 10^{-12}$ \\
$L_{\rm 2-10}$ & absorbed ($\ergps$) & $8.37 \times 10^{41}$ & $4.02
\times 10^{41}$ & $2.65 \times 10^{41}$ \\
$F_{\rm 2-10}$ & unabsorbed ($\ergpcmsqps$) & $2.25 \times 10^{-11}$ &
  $2.12 \times 10^{-11}$ & $1.59 \times 10^{-11}$ \\
$L_{\rm 2-10}$ & unabsorbed ($\ergps$) & $1.49 \times 10^{42}$ & $1.40
    \times 10^{42}$ & $1.05 \times 10^{42}$ \\
\hline\hline
Joint Fit & $\chi^2/\nu$ &  $3063/2668\,(1.15)$ & & \\
\hline\hline              
\end{tabular}
\end{center}
\caption{\small{Best-fit parameters and their errors (to $90\%$
    confidence for one interesting parameter) for the joint spectral
    modeling of the time-averaged datasets for the 2008, 2010a and 2010b
    observations. 
Only the XIS-FI and PIN data
    are included.  The energy range covered is
    $0.7-1.5$ and $2.5-40.0 \keV$, though only model components relevant for $E \geq 3
    \keV$ are listed here since the spectrum below this energy is dominated by
    extended, non-nuclear emission \citep{Wang2009}.
    Galactic column is fixed at $N_{\rm H}=1.34 \times 10^{20}
    \pcmsq$, as per \citet{Kalberla2005}.  The
    power-law indices of the {\tt reflionx} components were tied to
    that of the primary {\tt powerlaw} component, and the iron
    abundance of the warm absorber was tied to that of the two {\tt
    reflionx} components for consistency.  Unless otherwise specified,
    redshifts were fixed at the cosmological value for NGC~1365:
    $z=0.00547$.  Values marked with an asterisk (*) were tied between epochs while
    those marked with an (f) were fixed during the
    fit.}}
\label{tab:TA_tab}
\end{table}
\twocolumn

\subsection{Time-resolved Spectra}
\label{sec:TR}

\begin{figure}
\centerline{
\includegraphics[width=0.35\textwidth,angle=270]{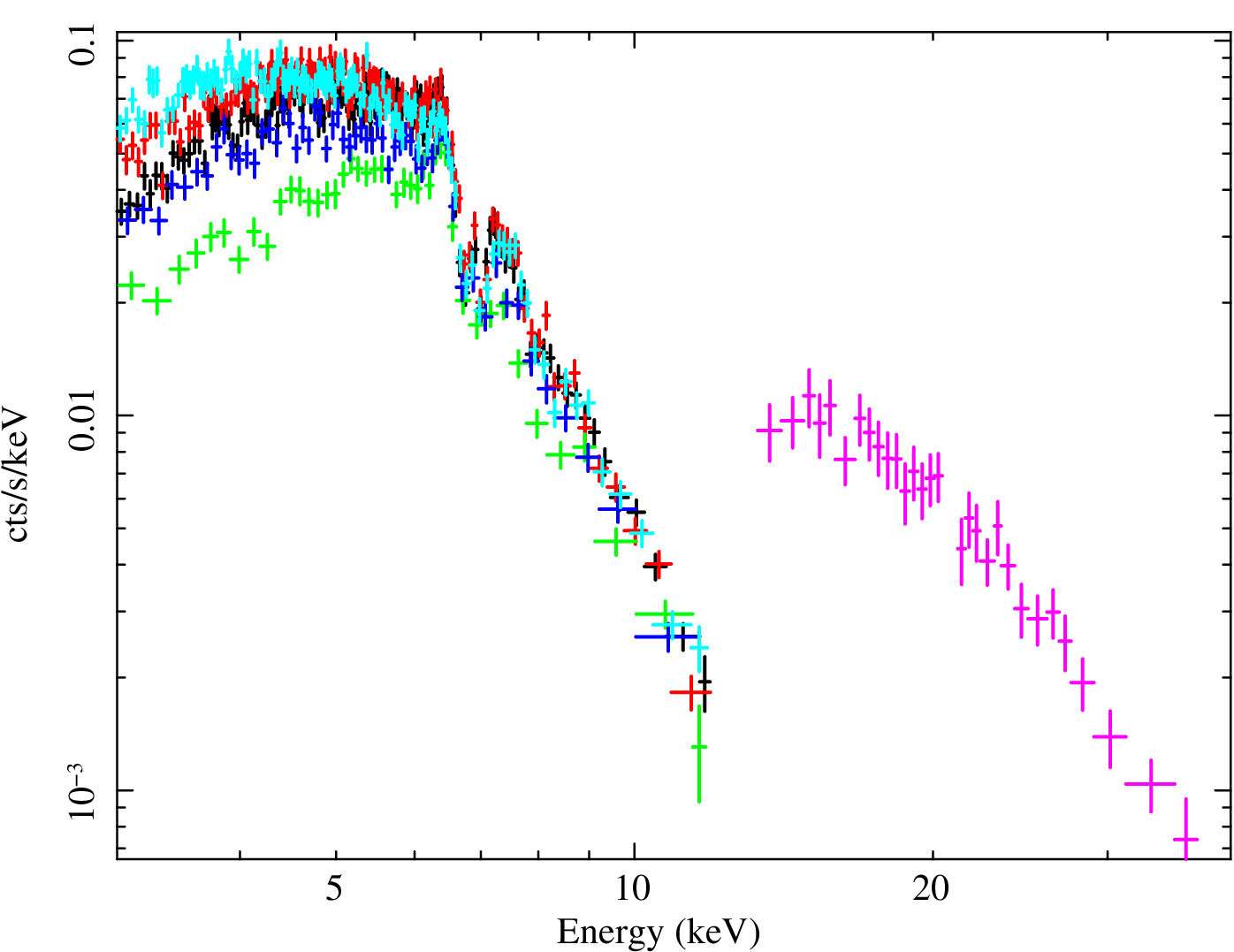}
}
\centerline{
\includegraphics[width=0.35\textwidth,angle=270]{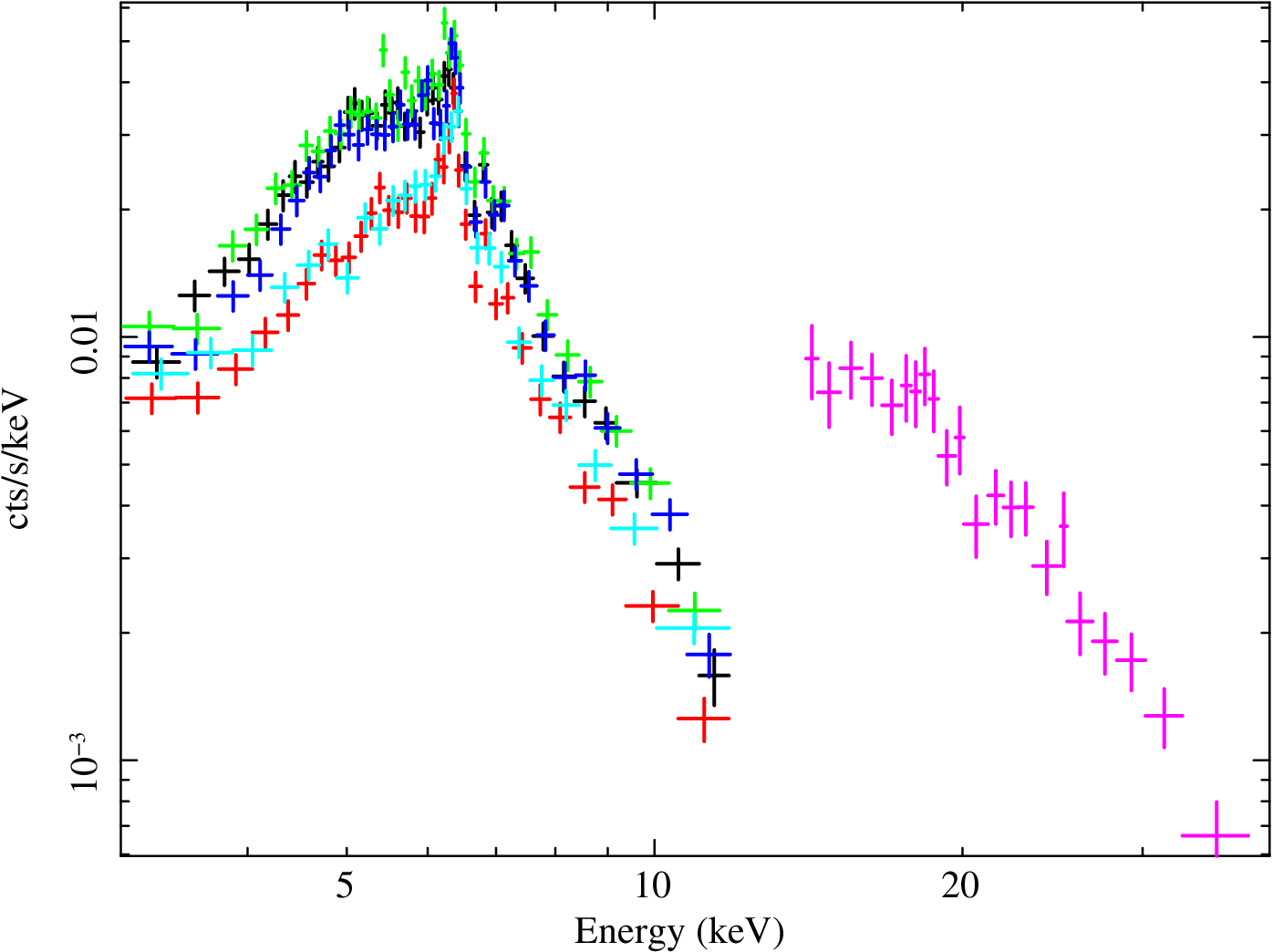}
}
\centerline{
\includegraphics[width=0.35\textwidth,angle=270]{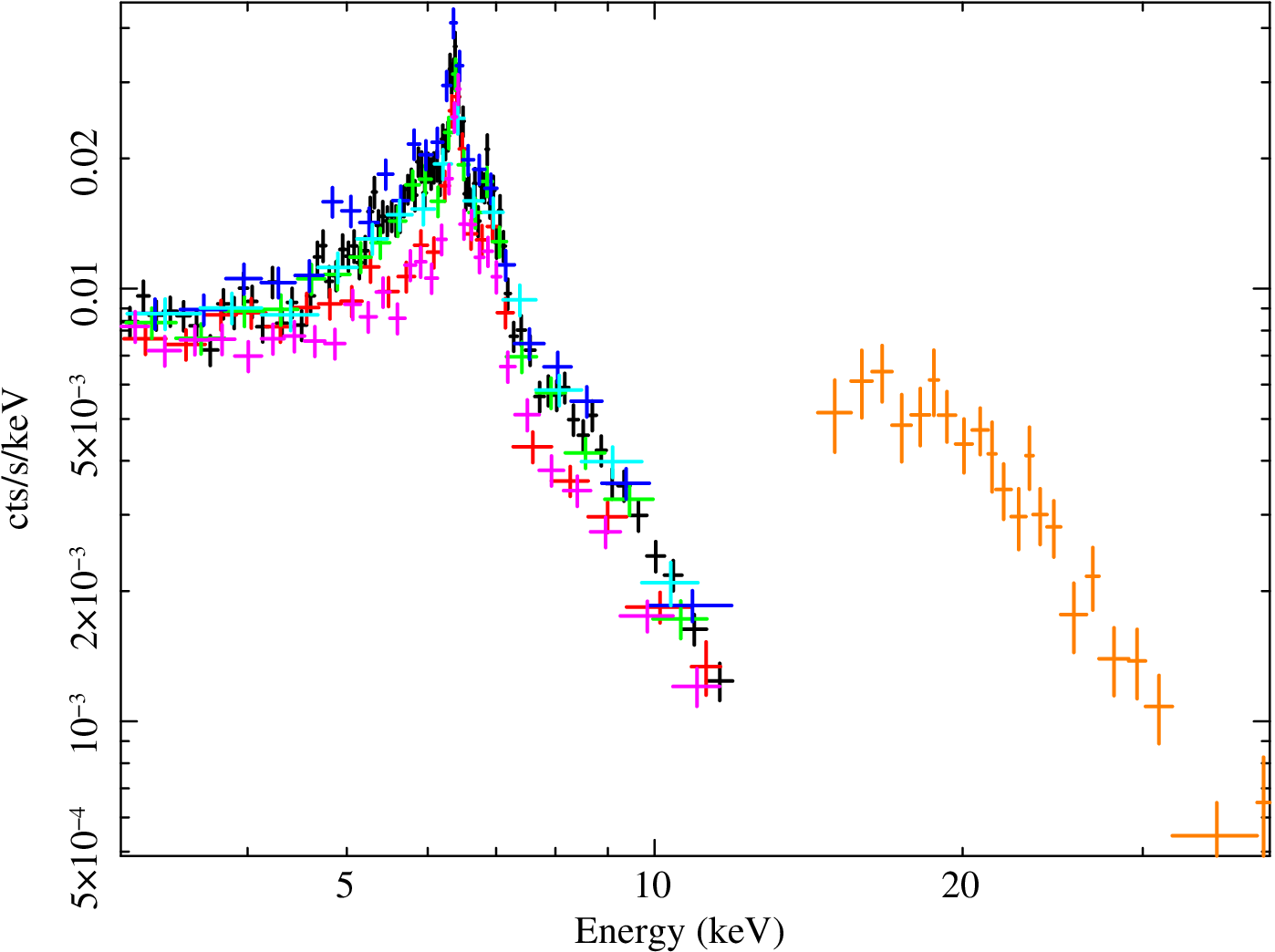}
}
\caption{{\small Time-resolved spectra from the five time intervals in 2008
    (top panel),
    2010a (center panel), and the six time intervals in 2010b (bottom panel).  Spectra
    shown are XIS-FI only for viewing purposes, and extend only from $3-10 \keV$
to avoid contamination from the starburst emission below these energies.
Time-averaged PIN
data are shown for each observation.  XIS interval
numbers in each panel are color-coded chronologically: black=interval 1,
red=2, green=3, dark blue=4, light blue=5, magenta=6 (2010b only).  PIN data are
shown in magenta in 2008 and 2010a and in orange in 2010b.}}
\label{fig:TR_spec}
\end{figure}

The lengths of our three {\it Suzaku} pointings of NGC~1365 also enabled us to examine
the spectral changes of this AGN within each observation.  We divided each
dataset into five (2008, 2010a) or six (2010b) time intervals, which are shown in
Fig.~\ref{fig:lc_hr}.  These intervals were selected based primarily
on the flux changes within each observation, though prominent hardness ratio
changes also played a role in helping us identify time intervals in
the 2008 observation.  Our selected time intervals are designed to focus on
changes in the spectral shape of the source from one observation segment to another.
The spectra extracted from each time interval in each observation are shown in
Fig.~\ref{fig:TR_spec}.  We focus on the XIS-FI spectra above $3 \keV$
in order to avoid the energies dominated by the $\sim$constant circumnuclear starburst emission
below $\sim3 \keV$.  The low energies are still included in the fit,
but the starburst parameter values are fixed to those listed in
Table~\ref{tab:TA_tab}.  The time-averaged PIN spectra for each observation are also
shown, and are considered in our analysis with a statistical weighting for each time
interval corresponding to the length of the time interval as a fraction of the
whole observation length, along with a flux normalization corresponding to the
flux of the XIS data in each interval relative to the time-averaged XIS flux.

\begin{figure}
\centerline{
\includegraphics[width=0.35\textwidth,angle=270]{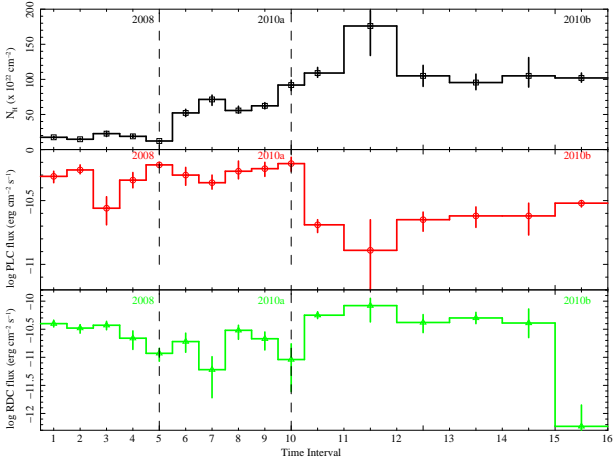}
}
\caption{{\small Variations in the cold absorbing column
    density (top panel), power-law
    flux (middle panel) and inner disk reflection flux
    (bottom panel) over the course of the three {\it Suzaku} observations of
    NGC~1365.  The 2008, 2010a and 2010b data are divided in each
    panel by the dashed vertical lines.  Time intervals are roughly
    twice as long for 2010b as for 2010a or 2008 since the observation
    length is roughly doubled.}}
\label{fig:var_plots}
\end{figure}

\begin{figure}
\centerline{
\includegraphics[width=0.35\textwidth,angle=270]{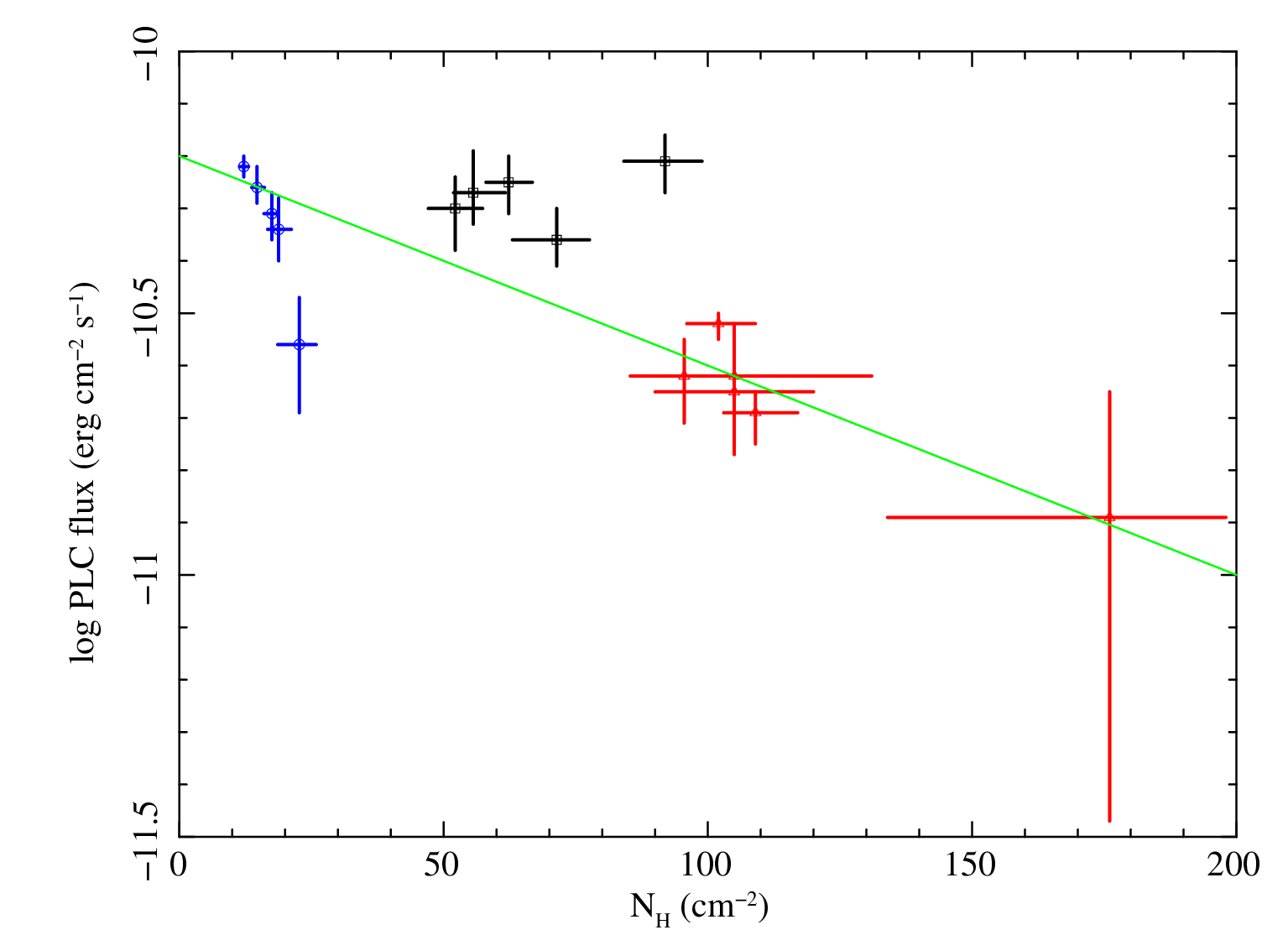}
}
\vspace{0.5cm}
\centerline{
\includegraphics[width=0.35\textwidth,angle=270]{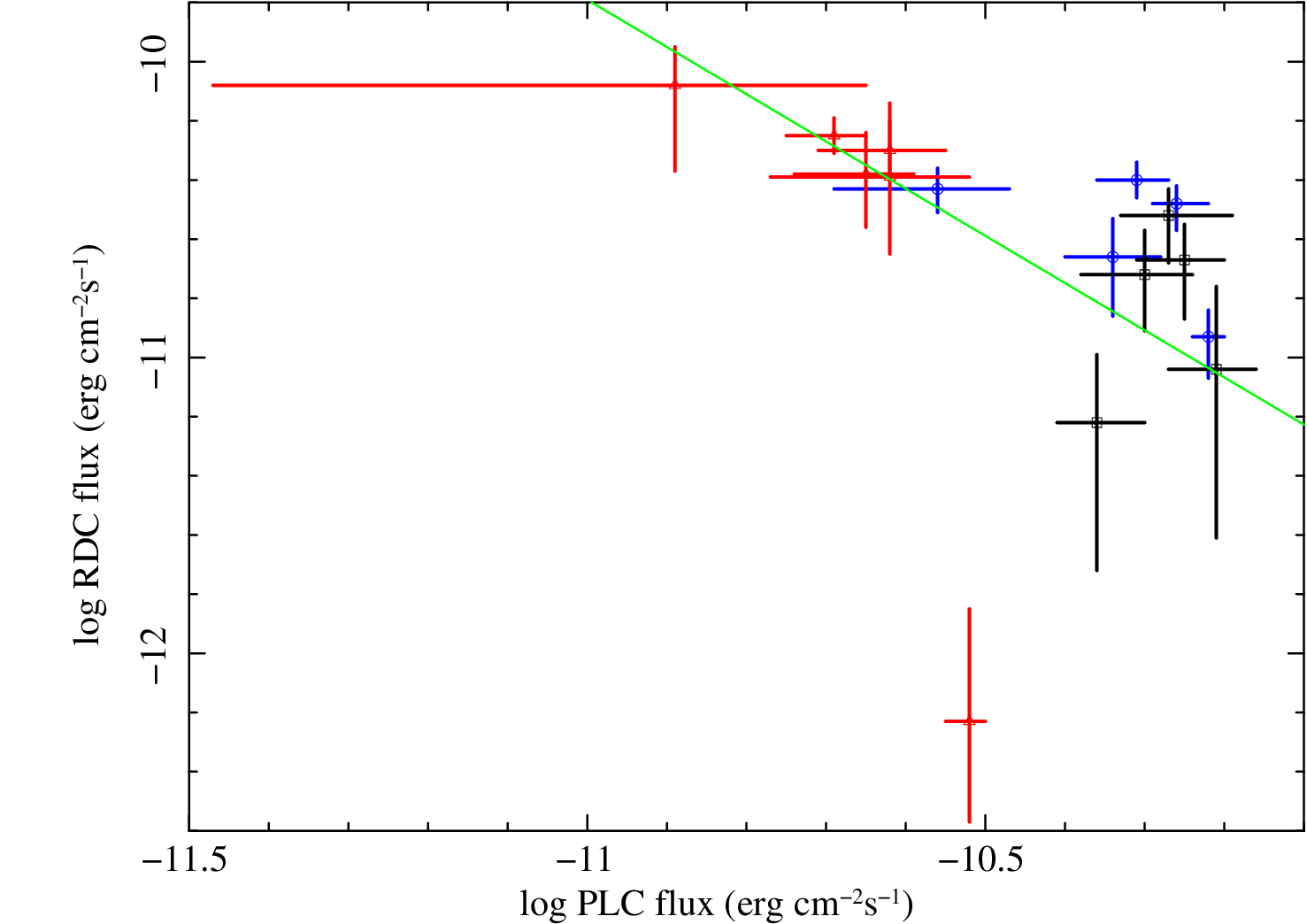}
}
\vspace{0.5cm}
\centerline{
\includegraphics[width=0.35\textwidth,angle=270]{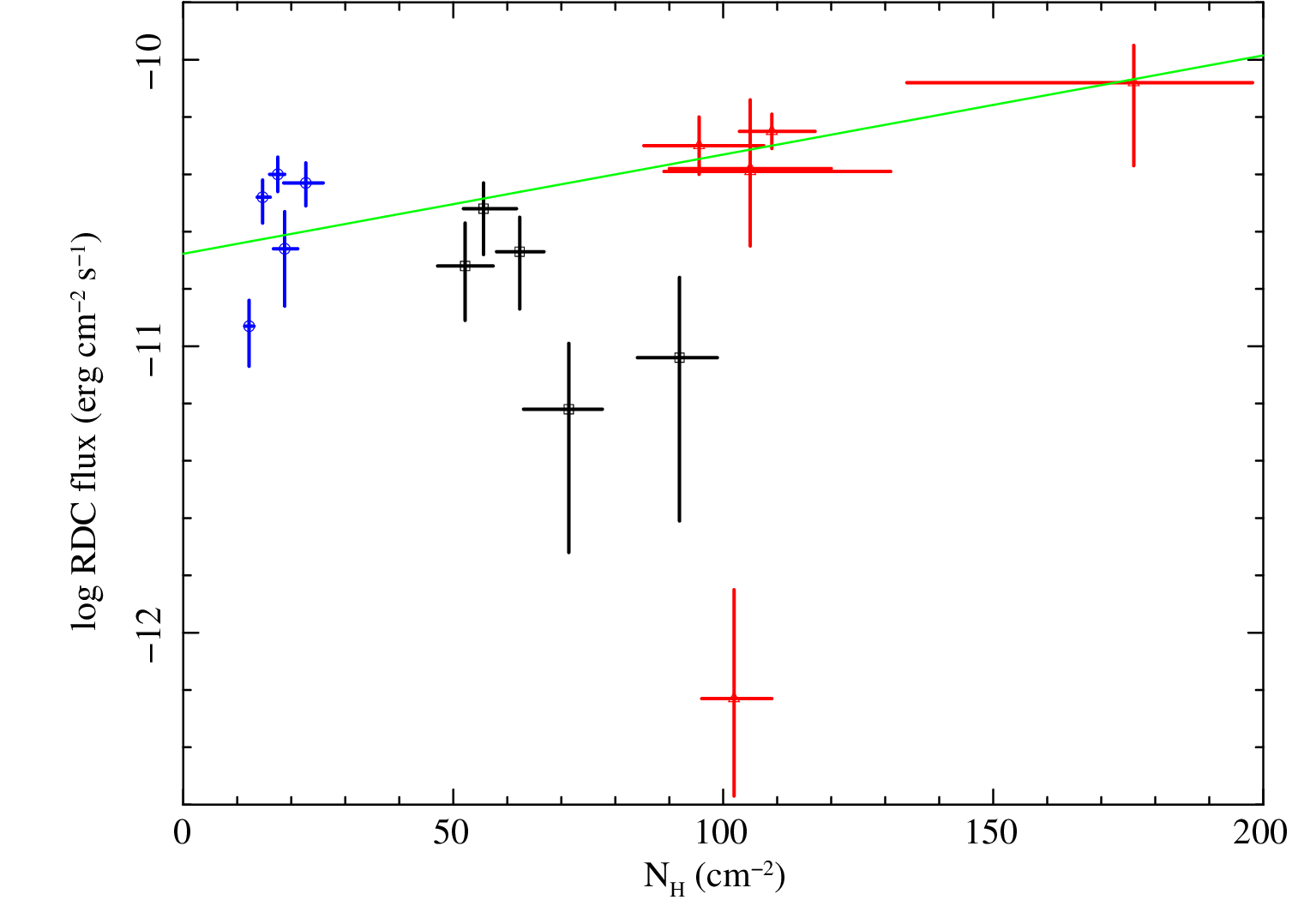}
}
\caption{{\small Plot of the variation in the cold absorbing column
    density (in units of $10^{22} \pcmsq$) vs. power-law component (PLC; top panel),
    the reflection-dominated component (RDC) vs. the PLC (middle
    panel), and the column density vs. the RDC (bottom panel) over
    the course of the three observations.  Data points are
    color-coded: blue circles for 2008, black squares for 2010a and
    red triangles for 2010b.  The solid green line represents a best
    fit to the data in each panel, neglecting those points with
    significantly larger errors than the norm.  The $N_{\rm H}$ vs. PLC
    plot does show an apparent inverse correlation, as does the PLC vs. RDC
    plot (though with a higher statistical uncertainty).  It then follows
    that the $N_{\rm H}$ vs. RDC plot shows a corresponding positive
    correlation, though also with a high statistical uncertainty.}}
\label{fig:correl}
\end{figure}

After fitting the time-averaged spectral model from each observation jointly to
its time intervals, and then allowing all of the parameters to vary between intervals, we
have determined that the only model components which vary
significantly (i.e., $\geq 3\sigma$) between
intervals are, not surprisingly, the cold absorbing column, the power-law
flux, and the flux of the inner disk reflector.  We therefore
proceed with our time-resolved spectral fitting for each observation allowing
only these three parameters to vary between time intervals, and fixing all other
parameters to their time-averaged values for that observation.  The results of
our time-resolved spectral analysis are presented in Table~\ref{tab:TR_tab}, and
are shown visually in Figs.~\ref{fig:var_plots}-\ref{fig:correl}.

The variation of the spectral shape between time intervals during each
observation is evident to the eye (Fig.~\ref{fig:TR_spec}).  An
examination of Table~\ref{tab:TR_tab}
and Figs.~\ref{fig:var_plots}-\ref{fig:correl} suggests that the cold absorbing column and
power-law component (PLC) flux are anti-correlated, as are the power-law flux and the flux
of the inner disk reflection-dominated component (RDC).  Closer inspection of the data
and errors in Fig.~\ref{fig:correl}
reveals that this apparent anti-correlation is most significant for
the PLC vs. $N_{\rm H}$ ($12\%$ uncertainty in line slope at $90\%$ confidence
due to scatter; Spearman's rank coefficient of $\rho=-0.703$ with
corresponding correlation P-value=0.001), but is less significant for the PLC vs. RDC ($27\%$
uncertainty; Spearman's rank coefficient of $\rho=-0.718$ with
corresponding correlation P-value=0.0008).
The RDC vs. $N_{\rm H}$ shows a positive correlation with $38\%$
uncertainty (Spearman's rank coefficient of $\rho=0.871$ with
corresponding correlation P-value=0.0001). 
 
The 2008 spectra are the brightest of the
three epochs and are also the least affected by the cold absorber, having a
convex shape below the Fe K$\alpha$ line (see Fig.~\ref{fig:TR_spec}).  The definitive
time-resolved spectral analysis of these data has been presented in
\citet{Maiolino2010}; as such, we do not repeat it unnecessarily, but
rather focus our examination on five more coarsely-spaced time
intervals to illustrate the overall changes in S/N during this observation.
Our time-resolved spectra show marked differences from
$3-6 \keV$, but above this energy they are nearly identical.  Joint spectral fitting of the five time
intervals demonstrates that the average $N_{\rm H} = (17.2^{+0.9}_{-1.0}) \times
10^{22} \pcmsq$, with a standard deviation of $\sigma=4.0 \times
10^{22} \pcmsq$ between
time intervals.  
The average power-law flux, log $F_{\rm PLC} =
-10.34^{+0.02}_{-0.03} \ergpcmsqps$, with $\sigma=0.13 \ergpcmsqps$
in log space between
intervals, and the average log $F_{\rm
  RDC} = -10.58^{+0.04}_{-0.06} \ergpcmsqps$ has $\sigma=0.22
\ergpcmsqps$ in log space between intervals.  

By contrast, the 2010a time-resolved spectra are slightly concave below $6
\keV$ (see Fig.~\ref{fig:TR_spec}), demonstrating that the absorbing
column is stronger during this observation
than in 2008.  The different 2010a time intervals do not show obvious
changes in
spectral shape, though they do show changes in amplitude, with intervals
$\#2$ and $\#5$ clearly having lower flux than intervals $\#1,\, \#3$ and
$\#4$ by a factor of $\sim1.5$ ($3\sigma$).  Joint spectral fitting of the five time
intervals in 2010a indicates that the average column density is $N_{\rm H} = (66.7^{+2.6}_{-2.8})
\times 10^{22} \pcmsq$, with a standard deviation of $\sigma=15.9
\times 10^{22} \pcmsq$ between intervals.  The average
power-law flux is log $F_{\rm PLC} = -10.28^{+0.03}_{-0.03}
\ergpcmsqps$, with $\sigma=0.06 \ergpcmsqps$ in log space between
intervals, and the average inner disk reflection
flux is log $F_{\rm RDC} = -10.83^{+0.08}_{-0.16} \ergpcmsqps$, with
$\sigma=0.29 \ergpcmsqps$ in log space between intervals.

The 2010b observation has the lowest flux of the three datasets and is also the
most heavily absorbed, as can be seen in its clear concavity below $6
\keV$ (see Fig.~\ref{fig:TA_spec_data}).
Whereas the 2008 and 2010a data showed clear evidence for changes in the cold
absorbing column both in their model fits and to the eye, the 2010b data do not
show such clear visual changes in absorption.  Overall, the spectral shape
change is less pronounced than in 2008 and 2010a, and is evenly split between
the absorbing column and the power-law flux, with no significant
contribution from the inner disk reflector.  On average, $N_{\rm H} =
(115^{+7}_{-8}) \times 10^{22} \pcmsq$, with a standard deviation of
$\sigma=30 \times 10^{22} \pcmsq$ between intervals.
The log $F_{\rm PLC} = -10.67^{+0.05}_{-0.10} \ergpcmsqps$, with
$\sigma=0.12 \ergpcmsqps$ in log space
between intervals.  The log $F_{\rm RDC} = -10.61^{+0.08}_{-0.09}
\ergpcmsqps$, with $\sigma=0.80 \ergpcmsqps$ in log space between
intervals.

\begin{figure}
\centerline{
\includegraphics[width=0.35\textwidth,angle=270]{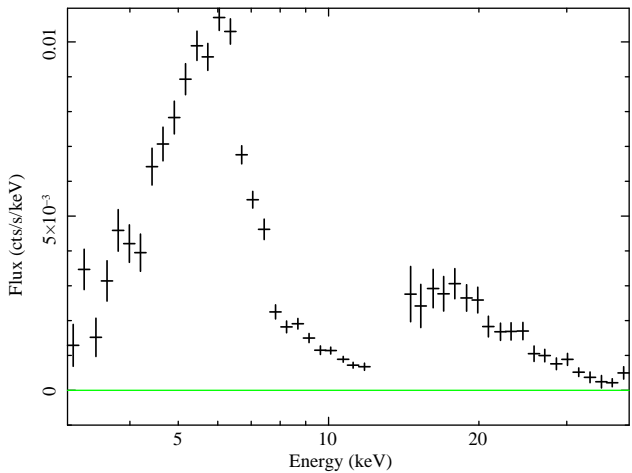}
}
\caption{{\small The co-added residual flux of the data above the
    model, left behind when the inner disk
    reflector is removed from the fit to all of the time-resolved
    datasets.  This plot depicts the features that the inner disk
    reflector accounts for in the spectral fit, and the statistical
    importance of those features.}}
\label{fig:rdc_tr_resid}
\end{figure}

We have demonstrated the importance of including relativistic, inner disk
reflection in the best-fit spectral model for the
time-averaged data in \S\ref{sec:TA}.  This component is equally
important in the spectral fitting of the time-resolved data.  To
illustrate this, Fig.~\ref{fig:rdc_tr_resid} shows the average residual left
behind when the inner disk reflection component (parametrized through
{\tt relconv[reflionx]}) is removed from the model in each time
  interval for each observation.  The residuals from each
  time-resolved interval are then co-added to produce the average
  residual shown.  Note the prominent, asymmetric Fe K$\alpha$ line
  and Compton hump that remain unmodeled without the inner disk
  reflection component.

In order to investigate the nature of the broad iron line and Compton
hump separately, we also attempted a model fit replacing the {\tt reflionx} component
for the inner disk with a {\tt pexrav} model plus a Gaussian line to represent
the Fe K$\alpha$ emission line core.  As with the {\tt reflionx} RDC,
we convolved the {\tt pexrav+zgauss} RDC with a {\tt relconv} smearing
kernel to account for relativistic effects near the black hole.  If
the broad line and Compton hump vary in the same sense (i.e., by
comparable magnitudes and on comparable timescales), this would
provide additional evidence supporting the relation and physical
origin of these two features in a way that {\tt reflionx} (which
assumes that they originate from the same material) cannot.
Unfortunately, neither the time-resolved spectra nor the time-averaged
spectra were able to constrain the parameters of both the {\tt pexrav} and
{\tt zgauss} components separately, mainly owing to the high
background of the PIN instrument.  {\it NuSTAR}
\citep{Harrison2005} will have both higher collecting area and a lower
background than the {\it Suzaku}/PIN instrument, however.  The {\it NuSTAR}
observing campaign on NGC~1365 is already underway and will enable
the broad iron line and Compton hump to be constrained separately and simultaneously
on these types of short timescales.

\onecolumn
\begin{table}
\begin{center}
\begin{tabular}{|lllll|}\hline\hline
{\bf Observation} & {\bf Interval} & {\bf Cold $N_{\rm H}\,(\times 10^{22}$)} &
{\bf log F$_{\rm PL}$ ($\ergpcmsqps$)} & {\bf log F$_{\rm Rrel}$ ($\ergpcmsqps$)} \\
\hline\hline
2008 & 1 & $17.5^{+1.3}_{-1.5}$ & $-10.31^{+0.04}_{-0.05}$ & $-10.40^{+0.06}_{-0.06}$ \\  
& 2 & $14.7^{+1.4}_{-1.0}$ & $-10.26^{+0.04}_{-0.03}$ & $-10.48^{+0.06}_{-0.09}$ \\  
& 3 & $22.7^{+3.2}_{-4.1}$ & $-10.56^{+0.09}_{-0.13}$ & $-10.43^{+0.07}_{-0.08}$ \\  
& 4 & $18.8^{+2.4}_{-2.1}$ & $-10.34^{+0.06}_{-0.06}$ & $-10.66^{+0.13}_{-0.20}$ \\  
& 5 & $12.2^{+0.9}_{-0.8}$ & $-10.22^{+0.02}_{-0.02}$ & $-10.93^{+0.09}_{-0.14}$ \\  
\hline
2010a & 1 & $52.2^{+5.2}_{-5.1}$ & $-10.30^{+0.06}_{-0.08}$ & $-10.72^{+0.15}_{-0.19}$ \\  
& 2 & $71.4^{+6.2}_{-8.4}$ & $-10.36^{+0.06}_{-0.05}$ & $-11.22^{+0.23}_{-0.50}$ \\  
& 3 & $55.6^{+6.1}_{-3.7}$ & $-10.27^{+0.08}_{-0.06}$ & $-10.52^{+0.09}_{-0.16}$ \\  
& 4 & $62.3^{+4.5}_{-4.3}$ & $-10.25^{+0.05}_{-0.06}$ & $-10.67^{+0.12}_{-0.20}$ \\  
& 5 & $91.9^{+7.0}_{-7.8}$ & $-10.21^{+0.05}_{-0.06}$ & $-11.04^{+0.28}_{-0.57}$ \\  
\hline
2010b & 1 & $109^{+8}_{-6}$ & $-10.69^{+0.04}_{-0.06}$ & $-10.25^{+0.06}_{-0.06}$ \\  
& 2 & $176^{+22}_{-42}$ & $-10.89^{+0.24}_{-0.58}$ & $-10.08^{+0.13}_{-0.29}$ \\  
& 3 & $105^{+15}_{-15}$ & $-10.65^{+0.06}_{-0.09}$ & $-10.38^{+0.14}_{-0.18}$ \\  
& 4 & $95.5^{+12.0}_{-10.2}$ & $-10.62^{+0.07}_{-0.09}$ & $-10.30^{+0.10}_{-0.10}$ \\  
& 5 & $105^{+26}_{-16}$ & $-10.62^{+0.10}_{-0.15}$ & $-10.39^{+0.25}_{-0.26}$ \\  
& 6 & $102^{+7 }_{-6}$ & $-10.52^{+0.02}_{-0.03}$ & $-12.23^{+0.38}_{-0.34}$ \\  
\hline
Joint Fit & $\chi^2/\nu$ & $4007/3795\,(1.06)$ & & \\
\hline\hline
\end{tabular}
\end{center}
\caption{{\small Parameter values and their errors (90\% confidence for one
    interesting parameter) for the best-fitting spectral models for the joint
    fits of the five time intervals in 2008 and 2010a and the six time intervals of 2010b,
    respectively.  The fit is over the energy range from $3-40 \keV$.  Absorbing
    column is in units of $\pcmsq$, and the log values of 
    power-law component (PLC) and inner disk reflection-dominated component (RDC)  
    flux are in units of $\ergpcmsqps$ as measured over the same
    energy band.}}
\label{tab:TR_tab}
\end{table}
\twocolumn 

\section{Discussion}
\label{sec:disc}

\subsection{Variability of the Cold Absorber}
\label{sec:disc_coldabs}

Historically, the eclipses of the inner accretion disk and corona (as tracked by
the broad Fe K$\alpha$ line and power-law continuum, respectively) in NGC~1365 have
been observed with a broad range of physical parameters.  The column
density of the occulting gas typically ranges from Compton-thin to
Compton-thick, with $N_{\rm H} \sim 2 \times 10^{23}-2 \times 10^{24} \pcmsq$.
The covering fraction of the occulting gas ranges from $\sim10\%$ to
  $\sim95\%$ between eclipse events, with one notable example of variation
  from $\sim55-100\%$, independent of column density variations,
  during the 2008 {\it Suzaku} observation of the source
  \citep{Maiolino2010}.
Such eclipses have been detected every time NGC~1365 has been observed
in X-rays with {\it Chandra, XMM} or {\it Suzaku}, and have typical durations of $40-70 \ks$
\citep{Risaliti2009c,Maiolino2010}.  Compton-thin eclipses are seen
more commonly than their Compton-thick counterparts, indicating that
the gas clumps along the line of sight to the central regions of the
AGN most often have $N_{\rm H} \leq 10^{24} \pcmsq$.

Our 2010 observations follow these trends, showing significant
variations in flux on timescales of days and weeks.  In both 2010a and 2010b,
we note two dips in the X-ray light curves (intervals $\#2$ and $\#5$
in 2010a and $\#2$ and $\#6$ in 2010b; see Fig.~\ref{fig:lc_hr}), though none of the dips is
greater than $25\%$ in flux (at $2\sigma$ significance), as compared with a $\geq40\%$ drop in
flux (at $10\sigma$ significance) seen in in the more dramatic 2008 eclipse events.  Further,
none of the flux decreases seen in 2010 corresponds to a significant
change in hardness ratio as seen in 2008.  The 2010a eclipses are all
Compton-thin, though all of the 2010b observation borders on a
Compton-thick state ($N_{\rm H} \sim 10^{24} \pcmsq$).  We also note
that the continued increase in column density of the cold absorber from
2008 through 2010b is a primary contributor to the overall decrease in
flux of NGC~1365 during this time.  Interestingly, the variability of
the source flux has decreased in accordance with the flux itself,
which might imply that the absorbing medium is more
uniform in addition to having a greater column density during 2010b
than 2010a or 2008 (but see \citealt{Uttley2001}: this direct
correlation between source flux and variability is also predicted in
systems where the flux is dominated by the coronal power-law
component, and the corona is flaring, or in scenarios where the
mass accretion rate in the disk is fluctuating on short timescales).  Though the
density argument is borne out by our spectral fitting, however, the
covering fraction of the cold gas in 2010b is actually smaller than in
2010a and 2008 (see Table~\ref{tab:TA_tab}).  

The low, relatively constant spectral state witnessed over $\sim300
\ks$ of on-source observing time in
2010b is a factor $\geq4$ longer than any previously observed eclipse events,
and may imply a new population of larger and/or slower moving cold absorber
clouds.  One
must also acknowledge the role of the diminished power-law flux during
this observation as well, however.  It is interesting to note the change in the
Fe K emission lines from 2008-2010: the narrow line core appears much more
prominent during 2010a and peaks in 2010b, while being virtually
absent in 2008 (see Fig.~\ref{fig:TA_spec_data}).  The normalization of the
distant reflector remains constant during this time.  This coincides with the
steady increase in column density of the cold absorber, a
diminishing of the broad red wing of the Fe K$\alpha$ line arising
from the inner accretion disk, and a diminishing of the flux from the
power-law component over the same time frame.  

In previous work on NGC~1365 by
\citet{Risaliti2005a,Risaliti2005b,Risaliti2007,Risaliti2009a,Risaliti2009b,Risaliti2009c},
and \citet{Maiolino2010},
constraints were placed upon the size of the inner disk/corona region ($R
\leq 10^{14} \cm$, or $\leq 170\,r_{\rm g}$ for a black hole with $M=4 \times
10^6 \Msun$, the mean of the three mass estimates quoted in
\citealt{Risaliti2009b}),
the distance of the eclipsing clouds from the black hole ($d \geq 2
\times 10^{15} \cm$, or $\geq 3400\,r_{\rm g}$) and the physical
parameters of the clouds themselves ($n_c \sim 10^{10}-10^{11} \pcmcu$,
$v_c \geq 1000 \kmps$).  

We have performed the same calculations with our data, interpreting
the four dips in the X-ray light curves of the 2010 observations as
eclipses in order to compare the constraints we derive on the physical
parameters of the system with those of Risaliti \etal and Maiolino
\etal  Assuming that the inner disk extends down to the marginally
stable orbit for a maximally rotating, prograde black hole ($r_{\rm
  in}=1.24\,r_{\rm g}$), and estimating the average time for the onset of the
eclipse to take place at $\geq5\ks$ (see Fig.~\ref{fig:lc_hr}; roughly
five times longer than the onset time for the eclipses in 2008), the
inferred cloud velocity is $v_c \leq 1470 \kmps$.  If this velocity is
Keplerian, we can then calculate a distance of the clouds from the
black hole of $d_c \geq 2.5 \times 10^{16} \cm$.
Assuming an average
2010 eclipse duration of $\sim80 \ks$ from Fig.~\ref{fig:lc_hr}
(3-4 times longer than eclipses in 2008), we can
place an upper limit on the size of the hard X-ray source of $r_x \leq 1.2
\times 10^{13} \cm$.  Estimating the density of the occulting clouds
from the column densities in the low-flux time intervals 2010a and
2010b observations, we find that the allowed range corresponds to
$n_c=6 \times 10^{11}-1.5 \times 10^{12} \pcmcu$. 

In short, the longer onsets of the eclipses in 2010
compared with those of 2008 yield cloud velocities roughly consistent with, though
slightly higher than those found in previous work.  The longer
durations of the 2010 eclipses, in tandem with their higher column
densities, yield significantly different constraints on the other
physical parameters of the system, however.  We find a larger density
for the absorbing gas by an order of magnitude.  The distance of the
clouds from the black hole is also a factor of ten higher based on our
data and calculations.  Correspondingly, the upper limit on the size
of the inner disk/corona region is a factor of ten smaller than found
in previous work.   
Risaliti \etal and Maiolino \etal state that the most likely location of
these clouds is the BELR ($\leq 10^{16} \cm$, or $\leq 17,000\,r_{\rm
  g}$).  If the dips in the two 2010 light curves are indeed eclipse
events, the absorbing clouds may lie on the outskirts of the BELR, or
beyond, in 2010.  If these are not true eclipse events with defined
clouds moving into and out of the line of sight to NGC~1365, but
rather represent periodic thickenings in a prolonged absorbed state,
then we may be seeing a new type of structure associated with the cold
absorber in this AGN.  In either case, the nature of the occulting
medium has clearly changed considerably in either distance or physical
structure from 2008 to 2010.

We do note that the combination of higher average column
density and lower average power-law flux in the 2010 observations
compared with the 2008 observation could result in the dips we observe
in the 2010 X-ray light curves having less contrast from the mean than
the clear eclipse events seen in 2008.  Even if the same clouds from
2008 were to pass through our line of sight in 2010, the overall lower
flux, more highly absorbed state of NGC~1365 at this time would
prevent us from seeing the same sharp changes in the light
curves and hardness ratios seen in the 2008 data.  The depth and
contrast of the dips in 2010 alone are not enough to rule out a
similar eclipsing cloud structure to that of 2008; rather, the duration of the dips
and length of onset of the dips are the primary indicators of a change
in the nature and/or location of the absorbing gas.

\subsection{The Nature of the Inner Disk/Corona Region}
\label{sec:disc_disk_corona}

In the 2008, 2010a and 2010b {\it Suzaku} data, we observe the
flux of the PLC changing significantly on
hours and weeks-long timescales
(Tables~\ref{tab:TA_tab}-\ref{tab:TR_tab}).  The 2008 and 2010a observations
coincidentally caught NGC~1365 with only a $\sim10\%$ difference
($1.6\sigma$) in
their average continuum strengths, but the power-law flux dropped by a
factor of $\sim2.1$ ($5.8\sigma$) in the two weeks between 2010a and 2010b.
The power-law strength did not change significantly during either 2010
observation, but the 2008 data do show a sudden drop of factor
$\sim2.0$ ($2.3\sigma$) from interval $\#2-3$, followed by a near-full
recovery by interval $\#4$. 

We can then examine the flux of the inner disk reflector (RDC) during these
time frames in the context of the PLC flux variation.  Though the
uncertainties on our measurements of the RDC flux are larger, on
average, than those on the PLC flux, we stress that the RDC is
nonetheless a necessary and variable component of the best-fit model
(see \S\ref{sec:TA}).  We note the following trends (see Table~\ref{tab:TR_tab} and
Figs.~\ref{fig:var_plots}-\ref{fig:correl}):

(i) The PLC and RDC fluxes vary significantly during 2008, though
these variations appear uncorrelated.  By contrast, the PLC and RDC
fluxes do not vary significantly within either 2010 observation
(though the RDC does decrease sharply, albeit with large error, during
the last time interval of 2010b).

(ii) Though there is no evidence suggesting correlated variability
within each observation, when taken together, the three observations
indicate a significant anti-correlation between the PLC and RDC (see
Fig.~\ref{fig:correl}), with a Spearman's rank coefficient of $\rho=-0.718$ with
a corresponding correlation P-value=0.0008 using the t-distribution.

In qualitative terms, trend (ii) is consistent with the
light-bending model of \citet{Miniutti2004} and \citet{Miniutti2006}, in which a compact corona
(or a compact, active region within a larger corona)
is located at some height $h$ from the accretion disk along the spin
axis of the black hole.  Inverse Compton-scattered photons from this
corona get reflected back down onto the surface of the underlying
disk, irradiating it with X-rays from the power-law continuum.  The degree of
anisotropy of this radiation depends on the height of the corona, with
lower heights corresponding to greater degrees of anisotropy, or
light-bending due to General Relativity.  Within the
light-bending paradigm, variations in the strengths of the
power-law vs. inner disk reflector can be attributed to variations in
the height of the corona along the spin axis.  A coronal height of $h \leq
4\,r_{\rm g}$ would result in the continuum and inner disk reflection
component normalizations being positively correlated, and $h \geq
4\,r_{\rm g}$ would result in these variables being uncorrelated (though
the quantitative values depend on geometry, e.g., 
\citealt{Niedzwiecki2008}).

An examination of the theoretical predictions for the relation between
the PLC and RDC fluxes as the coronal intrinsic luminosity changes
is presented in Fig.~7c of \citet{Niedzwiecki2008}.  Here, the author
assumes that the black hole is maximally rotating in the prograde
sense with $a=0.998$, and that the changes in PLC flux correspond to
the intrinsic luminosity of the source changing with the 
distance of the primary emitting
region of the corona from the disk surface according to radial emissivity
of a Keplerian disk.  This predicted RDC vs. PLC flux relation bears a
striking resemblance to the middle panel of our Fig.~~\ref{fig:correl}, depicting the
observed relationship between these two variables in our spectral
fitting for NGC~1365.  Our data points most closely mirror the
restricted region of parameter space in the \citet{Niedzwiecki2008}
plot around
the inflection point, which corresponds to a coronal distance of $h \leq
6\,r_{\rm g}$ from the disk.  The similarity of our Fig.~~\ref{fig:correl} to the
theoretical prediction
of \citet{Niedzwiecki2008} supports our inferences of a compact corona
and high black hole spin in this source, where the changes in the PLC
flux are due not only to changes in the location of the emitting
region(s) of the corona, but also to changes in the intrinsic luminosity of the corona
between and within our observations.
%
%

We will discuss our results within the framework of the
light-bending model in greater depth in Brenneman \etal (2013, in
prep.).

\subsection{Variability of the Warm Absorber}
\label{sec:disc_warmabs}

A significant column of ionized absorbing gas has been observed in NGC~1365 during the
majority of the X-ray pointings to date, with column densities, ionizations and
outflow velocities ranging from $N_{\rm H} \leq 10^{22}-5 \times 10^{23}
\pcmsq$, $\xi \sim 3000-5500 \ergcmps$ and $v_{\rm out} \sim 1000-5000 \kmps$,
respectively.  This warm absorber manifests as the Fe\,{\sc xxv} and
Fe\,{\sc xxvi} K$\alpha$ and K$\beta$ absorption lines from $6.7-8.3
\keV$ \citep{Risaliti2005b,Risaliti2007}.  Though this
warm absorber has not shown variability within observations (i.e.,
hours- and days-long timescales),
significant variability in column density, ionization and outflow velocity
has been noted on timescales of weeks and longer.  

The three {\it Suzaku}
observations of NGC~1365 showcase the nature of this variability.  No
significant changes are seen in the warm absorber properties during
2008 ($\sim160 \ks$), 2010a ($\sim150 \ks$) or 2010b ($\sim300 \ks$),
but the column density, ionization and outflow velocity all vary
significantly between pointings.  From the two years between 2008 and 2010a, $\Delta N_{\rm
  H}$ changed by a factor of $+1.7^{+0.3}_{-0.3}$, $\Delta \xi$ changed by a
factor of $+1.8^{+0.4}_{-0.3}$, and $\Delta v_{\rm out}$ changed by a factor of
$-1.9^{+1.1}_{-0.4}$.  In the
two weeks between 2010a and 2010b the absorption lines nearly
disappear, with $\Delta N_{\rm
  H}$ changed by a factor of $-10.3^{+1.1}_{-4.2}$, $\Delta \xi$ changed by a
factor of $-1.8^{+0.6}_{-0.5}$, and $\Delta v_{\rm out}$ changed by a factor of
$+1.6^{+1.2}_{-0.9}$.  Note
  that the increase in outflow velocity from 2010a to 2010b is within
  the error bars of the two parameters and is therefore not
  statistically significant.  The
reason for the lower limit on the noticeably large drop in column density between 2010a
and 2010b is that the 2010b column density is at the lowest end of its
parameter space in our {\sc xstar} table model. 

The lower limit of $300 \ks$ and upper limit of $\sim2$ weeks for the
warm absorber to vary significantly yields constraints on its distance from the
source of the ionizing photons (presumed to be the corona) of $9
\times 10^{15} - 3.6 \times 10^{16} \cm$, or $\sim15,000-61,000\,r_{\rm
  g}$ for NGC~1365.  This is based solely on light-crossing time, and assumes
that the warm absorber gas responds instantaneously to changes in the ionizing
radiation source (which is almost certainly not true).  The power-law continuum produced by the corona
clearly varies over much shorter timescales, so the distance to the
absorbing gas must be comparatively large in order for the absorption
lines not to vary on correspondingly short timescales, provided that
the physical properties of the warm absorbing gas remain constant.

We can also place limits on the distance of the absorbing gas from the hard X-ray source using the
ionization and column density of the gas derived from our spectral fitting, as
well as the $2-10 \keV$ luminosity of the ionizing continuum
(Table~\ref{tab:TA_tab}).  This method is likely more reliable than
using the variability timescale of the absorber alone, given the
uncertainties and assumptions involved in that calculation.
Calculating the distance of the gas using its ionization parameter
yields distance estimates of $2.8-3.2 \times 10^{15}
\cm$ for 2008, $1.1-1.9 \times 10^{14} \cm$ for 2010a and $2.5-3.9 \times
10^{14} \cm$ for 2010b, assuming that the
absorber is in ionization equilibrium and that its thickness along the radial
direction is roughly equal to its distance from the hard X-ray source.  In other
words, the absorbing gas is located between $185-5400\,r_{\rm g}$ in the system
consistent with a wind launched from the accretion disk \citep{Czerny2012}. 

We note that the power-law normalization stays relatively constant from 2008
to 2010a, then decreases significantly in 2010b, while the column of cold
absorbing gas steadily increases
from 2008 through 2010b.  Neither component shows any apparent correlation
with the variability of the warm absorbing gas.

\subsection{Future Work: Mapping the Inner Accretion Disk}
\label{sec:disc_future}

Given the clear evidence for both inner accretion disk reflection signatures and
a variable cold absorbing medium along our line of sight to the inner
disk, NGC~1365 is one of only a few sources in which eclipses of the
inner disk by clumps of the cold absorber can be used to examine the structure
of the disk and the nature of its reflection features in detail.  As
described in detail by \citet{Risaliti2011b}, the differences between
spectra from finely-spaced time slices during an eclipse by a
Compton-thick cloud with an eclipsed/non-eclipsed column density ratio
of $N_{\rm H1}/N_{\rm H2} \geq 10$ can effectively
provide a map of the flux and emissivity from isolated, progressive chords of the
inner disk.  

\begin{figure}
\centerline{
\includegraphics[width=0.35\textwidth,angle=270]{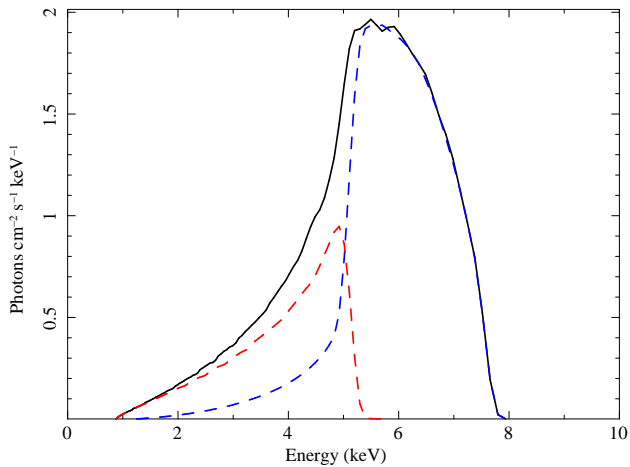}\\
}
\caption{{\small Expected change in the Fe K$\alpha$ line profile
    emitted from the inner accretion disk around a black hole with the
    parameters of NGC~1365 as described in Table~\ref{tab:TA_tab}.
    The solid black line shows the unocculted line profile, the blue
    (higher amplitude) dashed line
    shows the profile expected if the receding side of the disk is
    covered by a Compton-thick cloud, and the red (lower amplitude)
    dashed line shows the same
    profile expected if the approaching side of the disk is covered.}}
\label{fig:profiles}
\end{figure}

This type of accretion disk tomography can be used to
trace the change in the morphology of the putative broad Fe K$\alpha$ line
during an eclipse, testing both its origin as a reflection signature
(as opposed to an artifact of multiple complex absorbers, as in
\citealt{Miller2009}), and whether it varies as expected
within the framework of General Relativity.  Assuming that the
eclipsing clouds rotate with Keplerian velocities in the same
direction as the disk rotates, only the approaching (blueshifted) side of
the disk will be occulted at the onset of the eclipse, followed by
complete coverage, followed by only the receding (redshifted) side of the
disk being occulted, followed by a total uncovering of the disk.  The
expected broad Fe K$\alpha$ profiles for a black hole and inner disk
in each of these scenarios is shown in Fig.~\ref{fig:profiles}, with
its physical parameters taken from our time-averaged data for the {\it
  Suzaku} observations of NGC~1365.  As
discussed in \citet{Risaliti2011b}, these changes in the line profile
would be easily detectable even with current instruments, provided that
the cold absorbing gas has sufficient column density during the
eclipse compared to the uneclipsed state of the source.  Future
larger-area X-ray telescopes would also be able to
perform this experiment during Compton-thin eclipse events with less
contrast than the scenario described above.

We are currently constructing a library of theoretical simulations
describing the temporal and spectral variation of the broad Fe K$\alpha$ line profile from
the inner disk during eclipse events of varying physical properties.
Once completed, this library will be adapted as a table model within
{\sc xspec} and other software packages that can be fit to spectra and
light curves during eclipse events.  This model will be publicly
available, and will enable users to constrain the physical properties
of the disk and the eclipsing gas.  

Further observations are clearly needed in order to unlock the
potential of these eclipse events as probes of the inner environments
of AGN.  NGC~1365, in particular, has displayed a partial
Compton-thick eclipse during 4 out of 13 observations with {\it Chandra, XMM}
and {\it Suzaku}, most strikingly with
{\it Chandra} in 2006, in which the source changed from Compton-thin
to Compton-thick and back over the course of four days \citep{Risaliti2007}.
Unfortunately, none of the observations of the source in this state
was of sufficient quality to perform the type of accretion disk tomography we
have discussed in this Section.  Our goal is ultimately to observe a
full Compton-thick eclipse with enough S/N to perform a true
time-resolved spectral and temporal analysis of the event with our new
model.  Such an analysis would be the first of its kind, and would
serve as an important benchmark for future variability studies of AGN
with evidence for complex intrinsic absorption along the line of sight
to the inner disk/corona.

\section{Conclusions}

We have jointly analyzed three deep {\it Suzaku} observations of
NGC~1365 dating from 2008-2010 in order to examine the spectral and
temporal variability of the hard X-ray source, inner disk reflection and cold
absorber known to eclipse the inner nucleus in this AGN.  Our analysis
of the broad-band X-ray spectrum from $0.7-40 \keV$ and its changes over days-,
weeks- and years-long timescales has resulted in the following conclusions:

\begin{itemize}
\item{The 2008 observation shows the highest flux ($1.27 \times 10^{-11}
  \ergpcmsqps$) and the highest overall variability (factor $\sim3$) of the
  three datasets, with the intra-observation flux and variability both
  decreasing through 2010a ($6.13 \times 10^{-12} \ergpcmsqps$; factor
  $\sim1.8$) and 2010b
  ($4.05 \times 10^{-12} \ergpcmsqps$; factor $\sim1.3$).}  
\item{The source is at its softest in
  2008, becoming harder through 2010a and 2010b, and the hardness ratio shows
  the most variability in 2008 as well.}  
\item{Whereas the variability within each observation is dominated by
the soft energies (i.e., $\leq5 \keV$) in 2008, in 2010a and 2010b the
variability becomes increasingly dominated by energies above $\sim7 \keV$.}
\item{The basic spectral components observed in all three datasets are a
  constant circumnuclear starburst, a highly-ionized, outflowing warm
  absorber, a clumpy cold absorber along the line of sight to the nucleus, hard
  X-ray emission from the corona, $\sim$neutral reflection from distant material
in the outer disk and/or torus, and broadened, skewed reflection features from
the inner disk.}
\item{The spectral changes between observations are most evident in the XIS
  data, particularly from $2-6 \keV$ (factor $\sim8$).  The PIN data also display significant
  variability between observations (factor $\sim2$), though not to the degree present in the XIS
  data.}
\item{Incorporating both the relativistic reflection component and the
cold absorber is necessary in order to adequately model the spectrum
in all three epochs and flux states.}
\item{In keeping with recent work on other type-1 AGN (e.g., NGC~3783,
\citealt{Brenneman2011}) and previous work on NGC~1365
\citep{Walton2010}, we find ${\rm Fe/solar}=3.5^{+0.3}_{-0.1}$.  This
super-solar iron abundance can readily explain the so-called ``hard
excess'' noted in NGC~1365 in previous works (e.g.,
\citealt{Risaliti2009c}) without invoking an additional, ad hoc
absorber partially covering the source.}
\item{While the cold absorbing column density increases by a factor of $7$ from 2008 through 2010a
  and 2010b, the power-law normalization decreases by a factor of $2.3$.  Neither
  the distant reflection nor the
  relativistic reflection vary significantly between observations when taking errors into
  account (though some variability of the normalization of the inner
  disk reflector is suggested).} 
\item{The broad Fe K$\alpha$ line appears more prominent compared
  to the narrow line in 2008
  and steadily less so through 2010.  If this is the case, it is contrary to the behavior
  of the broad iron line in, e.g., MCG--6-30-15, in which the broad
  line becomes significantly more prominent as the power-law flux
  diminishes \citep{Fabian2002}.  An alternative explanation may be
  that the increase in cold absorbing column density from 2008 through
  2010 cuts into the red wing of the broad Fe K$\alpha$ line,
  diminishing the appearance of this feature.}  
\item{The warm absorber increases by roughly a factor of $2$ in column density
  and ionization in the two years between 2008 and 2010a, simultaneously
  dropping in outflow velocity by a factor of $2$.  But while the
  ionization drops and the outflow velocity rises again to its 2008 level in
  2010b, the column density plummets
by a factor of $\geq10$ in the two weeks between the 2010 observations.  This
variability, combined with ionization arguments, indicates that the
  warm absorbing gas must be located between 
$\sim185-60,000\,r_{\rm g}$ from the hard X-ray source
  within the BELR of
NGC~1365.  The variability of the WA is not correlated with that of
  the cold absorber.}
\item{Only three spectral components vary significantly within each
  observation: the power-law, inner disk reflector and cold absorber.
  The cold absorber shows the most statistically
  significant variations within each observation (typically
  $\sim25\%$).  The power-law varies comparably, though with slightly
  less statistical significance, and the inner disk reflector appears
  to vary by eye, though the errors on these measurements preclude a
  statistically sound interpretation of real variability within each
  observation.} 
\item{Within the framework of the light-bending model, we use these
  variations in the time-resolved data to infer
that the corona could be varying in
height from $h \geq 10\,r_{\rm g}$
during 2008 down to $h \sim 4\,r_{\rm g}$ in the
beginning of 2010a before rising to its original height through the
  beginning of 2010b, then
sinking back down to $h=4-10\,r_{\rm g}$ in the latter part of the
observation.  We
estimate that the lower limit for the timescale on which the corona varies significantly in
height is $\leq60 \ks$.}
\item{If we assume that the two dips in each 2010 observation of
  NGC~1365 correspond to eclipse events, we calculate that the
  eclipsing clouds must be a a distance $d_c \geq 2.5 \times 10^{16}
  \cm$, or $\geq 44,000\,r_{\rm g}$ from the black hole.  This is
  an order of magnitude more distant than the eclipsing clouds in the
  2008 observation.  Alternatively, if the dips represent a thickening
  of more of a continuous haze rather than discrete clouds eclipsing the
  inner disk and corona, we are seeing a new phase of the cold
  absorbing gas.  In either case, the nature of the cold absorber in
  2010 is significantly different from previous epochs.}
\item{An observation of a full, Compton-thick eclipse with a factor of
$\geq10$ change in column density could be captured with high enough
  S/N and spectral resolution with current instruments to perform
  accretion disk tomography.  Mapping the reflected X-ray emission from the
  disk in this way would definitively identify the origin of the
  putative broad iron line  in NGC~1365, and would provide valuable constraints
on the physical properties of the inner disk and eclipsing cloud(s).}
\end{itemize}

\bigskip \bigskip

\centerline{{\it Acknowledgments}}

We gratefully acknowledge support from the {\it Suzaku} mission
through NASA grant NNXlOAR44G, as well as the dedicated work of the
NASA/GSFC {\it Suzaku} GOF team on data calibration.  This work made
use of archived {\it Suzaku} data maintained by HEASARC at NASA/GSFC.
LB thanks Chris Reynolds and Mike Nowak for useful discussions that
contributed to the data analysis in this paper, and Dominic Walton for
providing the {\sc xstar} grid used in this work, as well as
insightful comments on the manuscript.  We also thank the anonymous
referee, whose critique greatly improved this work.

\bigskip


\bibliographystyle{mn2e_williams}
\bibliography{adsrefs}

\label{lastpage}

\end{document}